\documentclass[10pt, twocolumn, notitlepage, superscriptaddress, prl, longbibliography]{revtex4-2}

\usepackage{here}
\usepackage{amsmath}         
\usepackage{graphicx}
\usepackage{braket}         
 \usepackage{lettrine}
 \usepackage{color}
 \usepackage{times}
\usepackage{hhline}
\usepackage{tabularx}
\usepackage{calc}
\usepackage{soul}

\usepackage[colorlinks=true, linkcolor=blue, citecolor=blue]{hyperref}%

\begin{document}
             
\title{Non-Collinear Spin Current for Switching of Chiral Magnetic Textures} 

\author{Dongwook Go}
\email{d.go@fz-juelich.de}
\affiliation{Peter Gr\"unberg Institut and Institute for Advanced Simulation, Forschungszentrum J\"ulich and JARA, 52425 J\"ulich, Germany \looseness=-1}
\affiliation{Institute of Physics, Johannes Gutenberg University Mainz, 55099 Mainz, Germany}

\author{Moritz Sallermann}
\affiliation{Peter Gr\"unberg Institut and Institute for Advanced Simulation, Forschungszentrum J\"ulich and JARA, 52425 J\"ulich, Germany \looseness=-1}
\affiliation{Science Institute and Faculty of Physical Sciences, University of Iceland, VR-III, 107 Reykjavík, Iceland}

\author{Fabian R. Lux}
\affiliation{Institute of Physics, Johannes Gutenberg University Mainz, 55099 Mainz, Germany}

\author{Stefan Bl\"ugel}
\affiliation{Peter Gr\"unberg Institut and Institute for Advanced Simulation, Forschungszentrum J\"ulich and JARA, 52425 J\"ulich, Germany \looseness=-1}

\author{Olena Gomonay}
\affiliation{Institute of Physics, Johannes Gutenberg University Mainz, 55099 Mainz, Germany}

\author{Yuriy Mokrousov}
\email{y.mokrousov@fz-juelich.de}
\affiliation{Peter Gr\"unberg Institut and Institute for Advanced Simulation, Forschungszentrum J\"ulich and JARA, 52425 J\"ulich, Germany \looseness=-1}
\affiliation{Institute of Physics, Johannes Gutenberg University Mainz, 55099 Mainz, Germany}

\begin{abstract}
We propose a concept of non-collinear spin current, whose spin polarization varies in space even in non-magnetic crystals. While it is commonly assumed that the spin polarization of the spin Hall current is uniform, asymmetric local crystal potential generally allows the spin polarization to be non-collinear in space. Based on microscopic considerations we demonstrate that such non-collinear spin Hall currents can be observed for example in layered Kagome Mn$_3$X (X = Ge, Sn) compounds. 
Moreover, by referring to atomistic spin dynamics simulations we show that non-collinear spin currents can be used to switch the chiral spin texture of Mn$_3$X in a deterministic way even in the absence of an external magnetic field. 
Our theoretical prediction can be readily tested in experiments, which will open a novel route toward electric control of complex spin structures in non-collinear antiferromagnets.
\end{abstract}

\date{\today}                 
\maketitle

Recent studies have shown that antiferromagnets (AFMs) can take the role of ferromagnets in spintronics \cite{Jungwirth2016, Baltz2018} for their promising features such as the high frequency nature of eigenmodes, which offers a unique opportunity to study ultrafast dynamics at terahertz frequencies~\cite{Kittel1952, Ross2015, Gomonay2018}. Furthermore, the resilience to an external magnetic field and absence of stray fields make AFMs advantageous for increasing the memory density. However, these properties make manipulating magnetic order in AFMs extremely difficult. One of the major breakthroughs to an era of antiferromagnetic spintronics was a realization that magnetic moments in an AFM can be electrically controlled at an atomic scale due to ``locally'' asymmetric environment within each sublattice, even when the global inversion symmetry is present \cite{Zelezny2014, Wadley2016, Zelezny2017}. While early studies have focused on collinear AFMs such as CuMnAs \cite{Wadley2016} and Mn$_2$Au \cite{Bodnar2018, Meinert2018, Zhou2018}, in recent years, frustrated non-collinear (NC) AFMs such as Mn$_3$X (X=Sn, Ge) and Mn$_3$Ir attracted great deal of attention. Despite a vanishingly small net magnetic moment, they exhibit pronounced anomalous Hall effect \cite{Chen2014, Kubler2014, Nakatsuji2015, Nayak2016}, anomalous Nernst effect \cite{Ikhlas2017, Reichlova2019}, and magneto-optical Kerr effect \cite{Higo2018}, which is in clear contrast to a common wisdom of conventional  ferromagnets. These effects are driven by momentum-space Berry curvature originating from the chiral spin texture in real space \cite{Chen2014, Kuroda2017}. In particular, Mn$_3$Sn is identified as a magnetic Weyl semimetal \cite{Kuroda2017}, which exhibits magnetotransport phenomena of topological origin, e.g. chiral anomaly \cite{Nielsen1983, Armitage2018}. Here, change of the magnetic structure subsequently affects topology of the band structure, which provides an exciting platform to study an interplay of chiral magnetic texture and electron band topology \cite{Tsai2020}. 

However, complexity of magnetic interactions and excitations makes manipulation of the magnetic moments in NC AFMs even more challenging than in collinear AFMs. Nonetheless, Tsai \emph{et al.} succeeded in switching the magnetic state in polycrystalline Mn$_3$Sn/Pt heterostructures by using the spin Hall effect (SHE) of Pt under an external magnetic field \cite{Tsai2020}. Meanwhile, Takeuchi \emph{et al.} investigated epitaxial Mn$_3$Sn/Pt heterostructures and discovered a coherent rotation of the chiral spin texture in the Kagome plane \cite{Takeuchi2021}, which is induced by the spin injection whose polarization is perpendicular to the Kagome plane \cite{Gomonay2015, Dasgupta2022}. However, we note that Refs.~\cite{Tsai2020, Takeuchi2021} utilized a conventional spin Hall current to induce magnetic excitations in Mn$_3$Sn, which to some extent is analogous to a situation in heavy metal/ferromagnet bilayers.

\begin{figure}[b!]
\includegraphics[angle=0, width=0.47\textwidth]{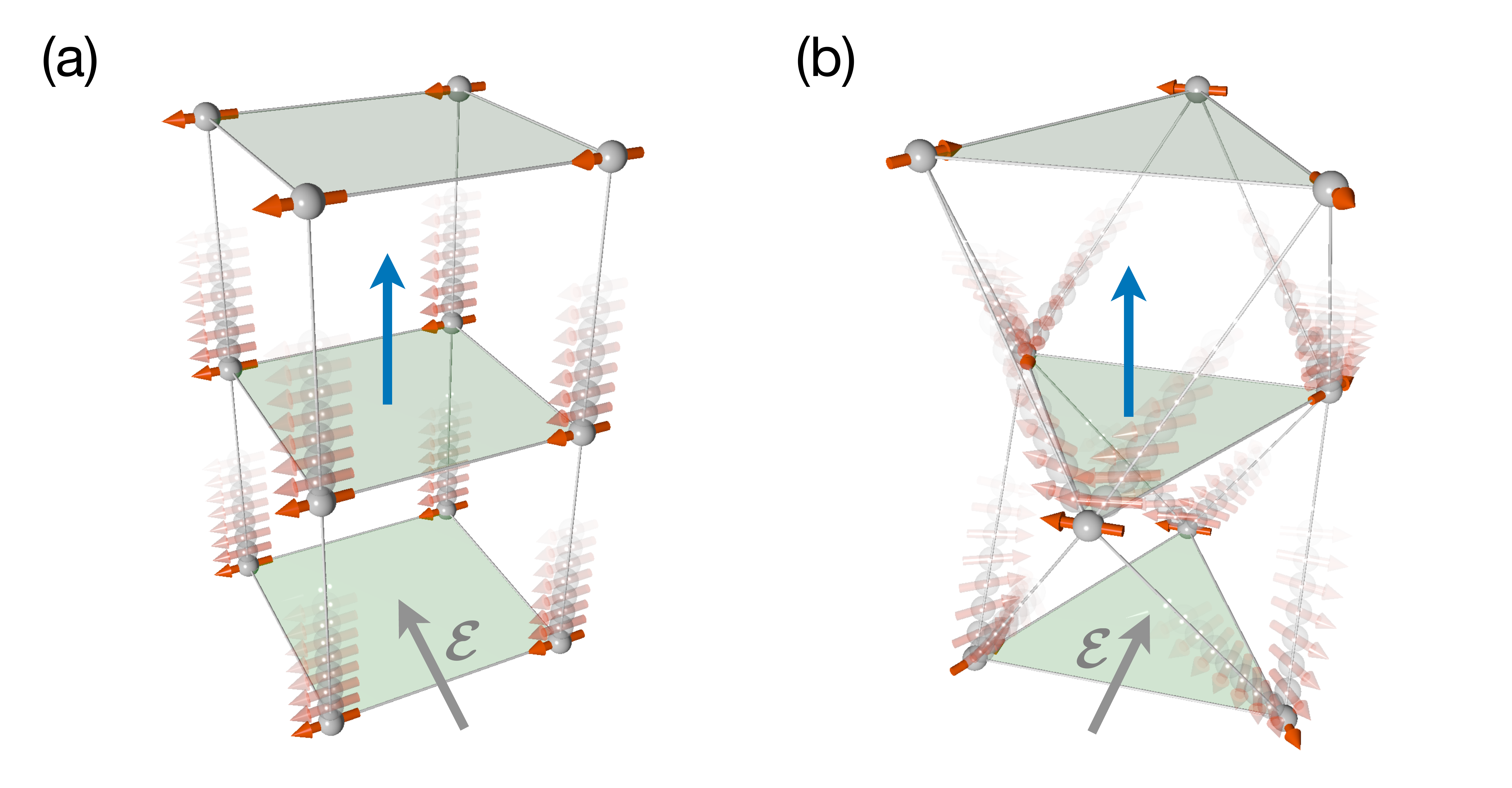}
\caption{
\label{fig:NC spin current}
Schematic illustration of the (a) conventional versus (b) NC spin Hall currents. The red arrows represent the direction of the spin polarization, the blue arrows indicate the electron's propagation direction in average, and the grey arrows represent the direction of an external electric field $\boldsymbol{\mathcal{E}}$.}
\end{figure}

A unique feature of AFMs is the sublattice degree of freedom, which allows for multiple modes of excitations. Keeping this in mind, we explore the possibility of a sublattice-dependent generation of spin Hall currents and investigate how they affect the dynamics of the magnetic texture in NC AFMs. While it is well-known that the spin polarization of the spin Hall current is orthogonal to both an external electric field and current propagation direction in cubic crystals without a sublattice degree of freedom \cite{Freimuth2010} [Fig.~\ref{fig:NC spin current}(a)], we find by symmetry arguments that the spin polarization of the spin Hall current can be NC in space already at the level of a non-magnetic crystal [Fig.~\ref{fig:NC spin current}(b)]. We demonstrate that such NC spin current can be generated in Mn$_3$X compounds, and it exhibits a ``chiral" component, which can be used to ignite a sublattice-dependent magnetization dynamics resulting in a switching of the magnetic texture. 
We believe that exploring the physics of sublattice-dependent NC spin currents would enrich our understanding of dynamics and excitations in various magnetic materials, and might provide a new way of efficient electric control of magnetic order.

First, we generalize the concept of a spin current to describe \emph{local} components in the vicinity of individual atoms, whose polarization may differ depending on the sublattice. The local spin current on site $i$ can be defined from the \emph{global} spin current $\mathbf{Q} = \mathbf{v}\otimes \mathbf{S}$ by 
$
\mathbf{Q}_i = (\mathbf{Q} P_i + P_i \mathbf{Q})/2,
$
where $\mathbf{v}$ is the velocity operator, $\mathbf{S}$ is the spin operator, and $P_i$ is the projection operator on site $i$, such that $\mathbf{Q} = \sum_i \mathbf{Q}_i$ \cite{Rauch2020}. Given an external electric field $\boldsymbol{\mathcal{E}}$, the local spin Hall conductivity (SHC) tensor on site $i$, $\sigma_{\alpha\beta,i}^{S_\gamma}$ is defined by $Q_{\alpha\beta,i} = \sigma_{\alpha\gamma,i}^{S_\beta} \mathcal{E}_\gamma$, where $\alpha,\beta,\gamma$ stand for Cartesian components of the velocity, spin, and external electric field, respectively. The \emph{global} SHC is recovered by summing the local SHC over the site index $i$: $\sigma_{\alpha\gamma}^{S_\beta} = \sum_i \sigma_{\alpha\gamma,i}^{S_\beta}$. We define the NC spin current as the local spin current whose spin polarization varies depending on atomic site $i$.

\begin{figure}[t!]
\includegraphics[angle=0, width=0.48\textwidth]{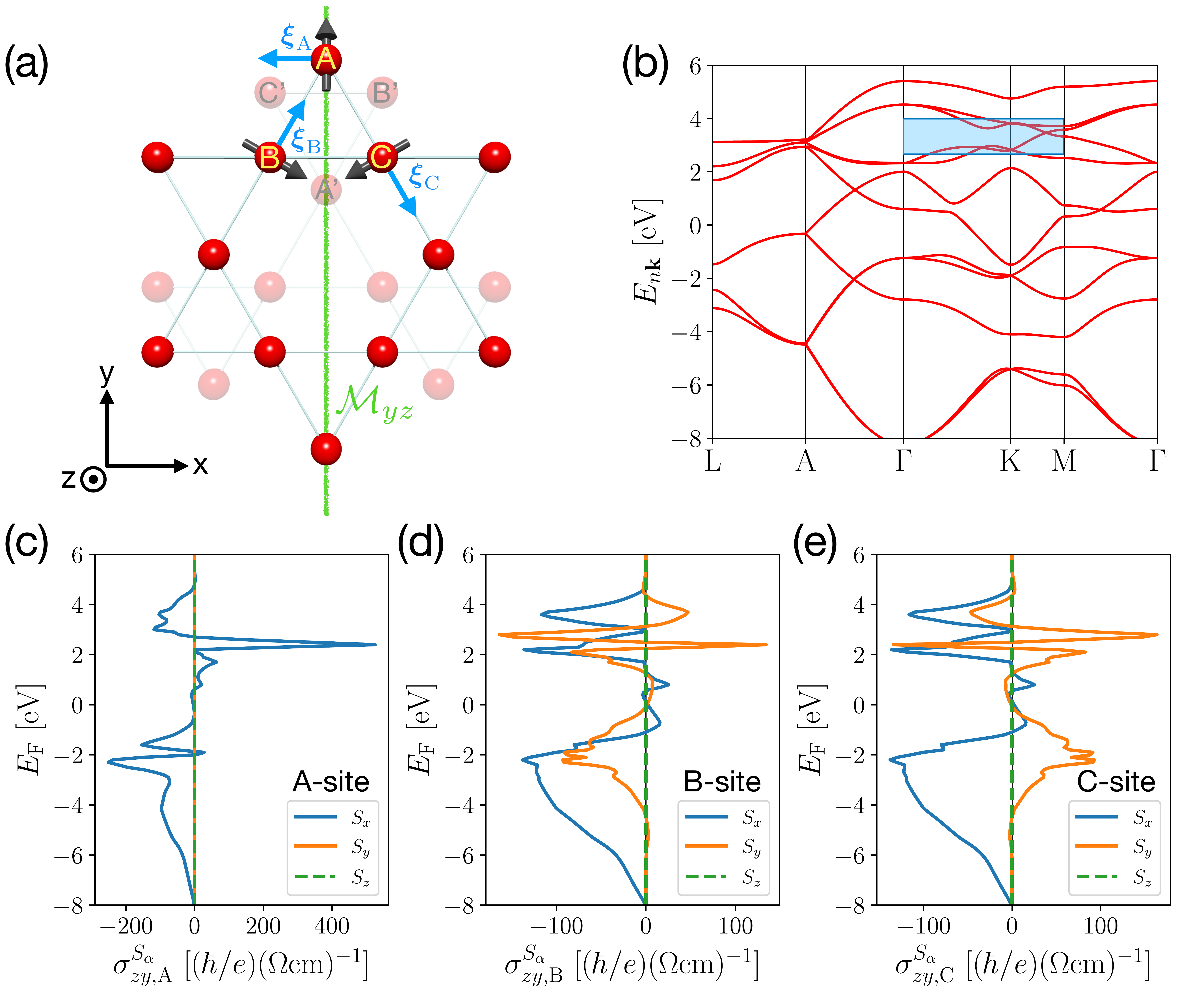}
\caption{
\label{fig:TB1}
(a) The structure of Mn$_3$X (X=Sn, Ge). Only Mn atom sites are shown as red spheres. There are 6 Mn atoms in a unit cell, and Mn atoms A, B, C in one layer are related to Mn atoms A$^\prime$, B$^\prime$, C$^\prime$ in another layer by the inversion symmetry. A mirror plane $\mathcal{M}_{yz}$ is indicated by a green line. The chiral directions [Eq.~\eqref{eq:chiral_SHC}] are represented by blue arrows. (b) Electronic band structure according to the TB model of MnX$_3$ compounds. Shown are the local SHCs on (c) A-site, (d) B-site, and (e) C-site, respectively.}
\end{figure}

A unique direction of the spin polarization for the \emph{global} spin current in the SHE is set by a mirror plane containing both an external electric field and electron's propagation path. This explains why an external electric field, electron's propagation, and the spin polarization are orthogonal to each other in cubic crystals. In Mn$_3$X, the  mirror plane $\mathcal{M}_{yz}$ [indicated by a green line in Fig.~\ref{fig:TB1}(a)] allows only for $S_x$ polarization of the global spin Hall current flowing along $z$ when an external electric field is applied along $y$. Because  $\mathcal{M}_{yz}$ transforms $Q_{zx}\rightarrow Q_{zx}$, $Q_{zy}\rightarrow -Q_{zy}$, and $Q_{zz}\rightarrow -Q_{zz}$ while $\mathcal{E}_y$ remains invariant, $\sigma_{zy}^{S_y}$ and $\sigma_{zy}^{S_z}$ are not allowed.

In Mn$_3$X, however, not all sublattice atoms are located on top of $\mathcal{M}_{yz}$ [Fig.~\ref{fig:TB1}(a)], which implies that the local spin Hall current can generally have NC polarization depending on the sublattice. Let us consider again a situation where an external electric field is applied along $y$ and the spin Hall current is flowing along $z$, and analyze the structure of the local SHC tensor, $\sigma_{zy,i}^{S_\beta}$. On site $\mathrm{A}$, only $S_x$ polarization is allowed because the site index $\mathrm{A}$ is invariant with respect to $\mathcal{M}_{yz}$. However, on sites $\mathrm{B}$ and $\mathrm{C}$, the spin current can have $S_y$ polarization, which is a nontrivial direction, as well as $S_x$ polarization. In general, there is no symmetry element that enforces them to vanish. Because site indices $\mathrm{B}$ and $\mathrm{C}$ are interchanged by $\mathcal{M}_{yz}$, these components are related by $\sigma_{zy,\mathrm{B}}^{S_y} = -\sigma_{zy,\mathrm{C}}^{S_y}$ and $\sigma_{zy,\mathrm{B}}^{S_x} = \sigma_{zy,\mathrm{C}}^{S_x}$. We remark that the NC spin current is ``hidden'' in this case: If the local SHCs are summed over all sites, only $\sigma_{zy}^{S_x}$ survives and $\sigma_{zy}^{S_y}$ vanishes. However, we remark that breaking of a mirror symmetry can induce occupation asymmetry of the hidden components, leading to an unconventional spin polarization of the global SHE \cite{Roy2021}. Meanwhile, the components of the local SHC on sites A$^\prime$, B$^\prime$, C$^\prime$ are identical to those on sites A, B, C by the inversion symmetry. 

In the following, we discuss only the time-reversal-\emph{even} component of the SHE. This means that the NC spin current we describe here arises even at the level of nonmagnetic crystals due to anisotropic crystal potential in combination with the spin-orbit coupling (SOC). This is different from the magnetic or time-reversal-\emph{odd} SHE \cite{Zelezny2017b, Kimata2019, Mook2020}, which arises from the magnetic texture. We emphasize that, generally, the interplay of the local symmetry and magnetic order leads to a very rich structure of the NC spin current since the local SHC depends not only on the site index but also on the direction of local magnetic moments. We leave an investigation of such higher-order ``magnetic" contributions to the NC spin current for future work.

For the demonstration of the emergence of NC spin current in Mn$_3$X, we adopt a tight-binding (TB) model. As shown in Fig.~\ref{fig:TB1}(a), the unit cell contains six atoms with $s$-like orbitals on each Mn site. The electronic Hamiltonian is written as
\begin{eqnarray}
\mathcal{H}_\mathrm{el} &= & -t \sum_{\langle ij \rangle}\sum_\alpha c_{i\alpha}^\dagger c_{j\alpha} 
+
J \sum_{i}\sum_{\alpha\beta} 
c_{i\alpha}^\dagger
\left(
\boldsymbol{\sigma}_{\alpha\beta}
\cdot
\hat{\mathbf{m}}_i
\right)
c_{i\beta}
\nonumber
\\
& & +
i\lambda \sum_{\langle ij \rangle} \sum_{\alpha\beta}
c_{i\alpha}^\dagger
\boldsymbol{\sigma}_{\alpha\beta}
\cdot 
\hat{\mathbf{n}}_{ij}
c_{j\beta},
\label{eq:Hamiltonian}
\end{eqnarray}
where the first, second, and third terms describe inter-site hopping, $sd$ exchange coupling, and SOC. For site indices $i,j$, the notation $\langle \cdots \rangle$ in the summation means that only  nearest neighbors are taken into account. Here, $c_{i\alpha} (c_{i\alpha}^\dagger)$ is an electron annihilation (creation) operator on site $i$ with spin $\alpha$ and $\boldsymbol{\sigma}_{\alpha\beta}$ is a matrix element of the vector of Pauli matrices. The parameters are set as follows: $t=1.0\ \mathrm{eV}$ for the nearest neighbor hopping amplitude, $J=1.7\ \mathrm{eV}$ for the $sd$ exchange interaction with local moment $\hat{\mathbf{m}}_i$, $\lambda=0.2\ \mathrm{eV}$ for the strength of the SOC. Meanwhile, $\hat{\mathbf{n}}_{ij}$ is a unit vector orthogonal to both the hopping direction and the local crystal field. We set the direction of the magnetic moments by $\hat{\mathbf{m}}_\mathrm{A}=\hat{\mathbf{y}}$, $\hat{\mathbf{m}}_\mathrm{B}=(\sqrt{3}/2)\hat{\mathbf{x}}-(1/2)\hat{\mathbf{y}}$, and $\hat{\mathbf{m}}_\mathrm{C}=-(\sqrt{3}/2)\hat{\mathbf{x}}-(1/2)\hat{\mathbf{y}}$, as shown by black arrows in Fig.~\ref{fig:TB1}(a). 

The electronic band structure of the TB model is shown in Fig.~\ref{fig:TB1}(b), which agrees with the result from the previous study \cite{Ito2017}. We evaluate the intrinsic \emph{local} SHC from the TB model for the spin current flowing along $z$ when an external electric field is applied along $y$. It is given by
\begin{eqnarray}
\sigma_{zy,i}^{S_\alpha}
=
\frac{e}{\hbar}
\sum_{n}
\int \frac{d^3k}{(2\pi)^3}
f_{n\mathbf{k}} \Omega_{n\mathbf{k},zy,i}^{S_\alpha},
\end{eqnarray}
where 
\begin{eqnarray}
& &
\Omega_{n\mathbf{k},zy,i}^{S_\alpha} 
=
2\hbar^2
\nonumber
\\
& &
\ \ \ \times \sum_{m \neq n}
\mathrm{Im}
\left[
\frac{
\bra{u_{n\mathbf{k}}}
Q_{z\alpha,i} 
\ket{u_{m\mathbf{k}}}
\bra{u_{m\mathbf{k}}}
v_y
\ket{u_{n\mathbf{k}}}
}
{(E_{n\mathbf{k}} - E_{m\mathbf{k}} + i\eta)^2}
\right]
\end{eqnarray}
is the atom-projected spin Berry curvature. Here, $e>0$ is the magnitude of the electron's charge, $\hbar$ is the reduced Planck constant, $u_{n\mathbf{k}}$ is periodic part of the Bloch state, and $E_{n\mathbf{k}}$ and $f_{n\mathbf{k}}$ are corresponding energy eigenvalue and Fermi-Dirac distribution, respectively. The calculated local SHC's on site A, B, C are shown in Figs.~\ref{fig:TB1}(c), \ref{fig:TB1}(d), \ref{fig:TB1}(e), respectively, as a function of the Fermi energy $E_\mathrm{F}$. As explained, the local SHC on site A has only $S_x$ component and the other spin components are absent. On site B and C, however, both $S_x$ and $S_y$ are present. The microscopic TB calculation clearly demonstrates the NC feature in the local spin current, whose spin polarization varies from site to site.

Considering a combined result of various components of the local conductivity tensor $\sigma_{zy,i}^{S_\alpha}$, we can
define the ``chiral'' component of the SHC by
\begin{eqnarray}
\sigma_{zy}^{S_\mathrm{ch}}
=
\sum_{i,\alpha}
\sigma_{zy,i}^{S_\alpha} \xi_{\alpha,i},
\label{eq:chiral_SHC}
\end{eqnarray}
where $
\hat{\boldsymbol{\xi}}_\mathrm{A}
=
-\hat{\mathbf{x}},
\
\hat{\boldsymbol{\xi}}_\mathrm{B}
=
(1/2)\hat{\mathbf{x}}
+(\sqrt{3}/{2}) \hat{\mathbf{y}}
,
\
\hat{\boldsymbol{\xi}}_\mathrm{C}
=
(1/2)\hat{\mathbf{x}}
-(\sqrt{3}/{2}) \hat{\mathbf{y}},
$ determine the ``chiral'' directions on each site, which are indicated by blue arrows in Fig.~\ref{fig:TB1}(a). This is the component that enables coupling to the NC texture of magnetic moments in Mn$_3$X and inducing a NC spin torque, which goes beyond the uniform and staggered torques conventionally discussed in the context of ferromagnetic or N\'eel order switching.

\begin{figure}[t!]
\includegraphics[angle=0, width=0.49\textwidth]{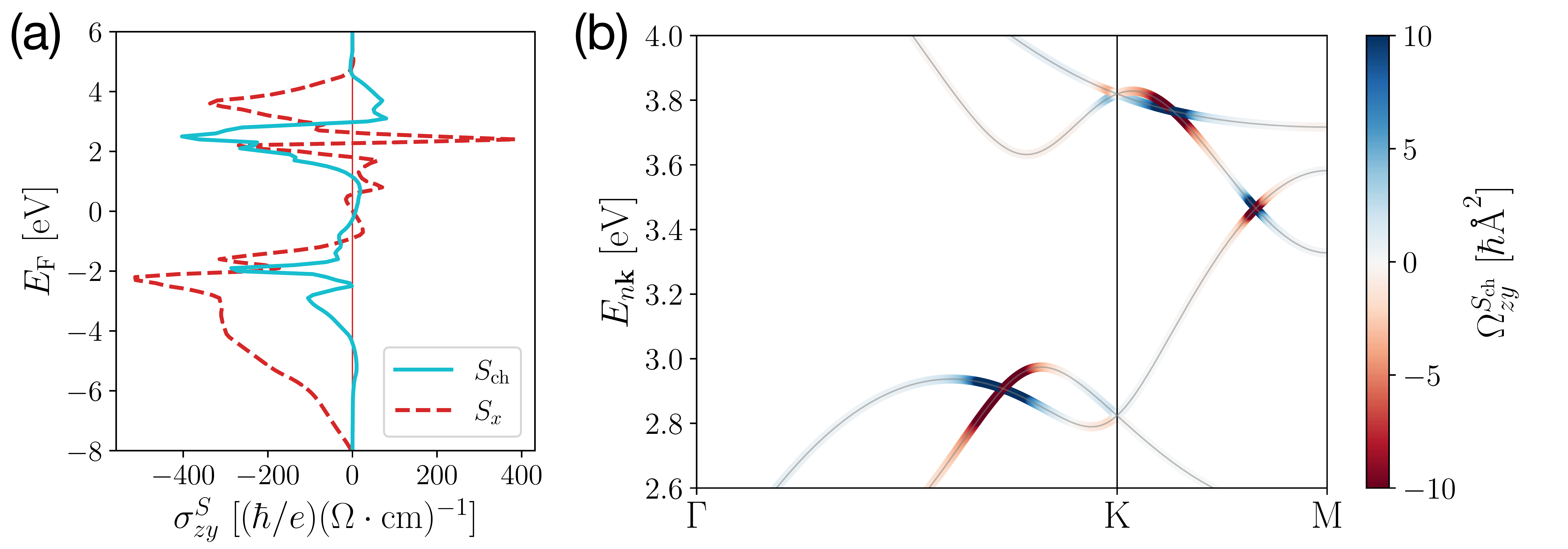}
\caption{
\label{fig:TB2}
(a) Comparison of the SHCs for the chiral component $S_\mathrm{ch}$ (cyan solid line) and a uniform component $S_x$ (red dashed line) as a function of the Fermi energy. (b) ``chiral'' spin Berry curvatures near the band crossings, which is indicated by color on top of the band structure.
}
\end{figure}

Our calculation of the chiral SHC is shown in Fig.~\ref{fig:TB2}(a), which is compared with a uniform ($S_x$) component. We observe that while the uniform component has large values over a wide range of energy, the chiral component tends to exhibit a more spiky behavior. This is because it requires a ``chiral" mixing of the spin character which can be achieved at specific $\mathbf{k}$-points in the electronic structure. However, it is remarkable that the chiral component can be as large as the uniform component, especially near $E_\mathrm{F}\approx -2\ \mathrm{eV}$ and $E_\mathrm{F}\approx +3\ \mathrm{eV}$ [highlighted by a blue box in Fig.~\ref{fig:TB1}(b)], where various bands cross each other. To visualize the influence of band crossings, in Figs.~\ref{fig:TB2}(b), we plot the ``chiral'' spin Berry curvature near $E\approx +3\ \mathrm{eV}$ along $\Gamma - \mathrm{K} - \mathrm{M}$, which is strongly pronounced near the band crossings. 

\begin{figure}[t!]
\includegraphics[angle=0, width=0.45\textwidth]{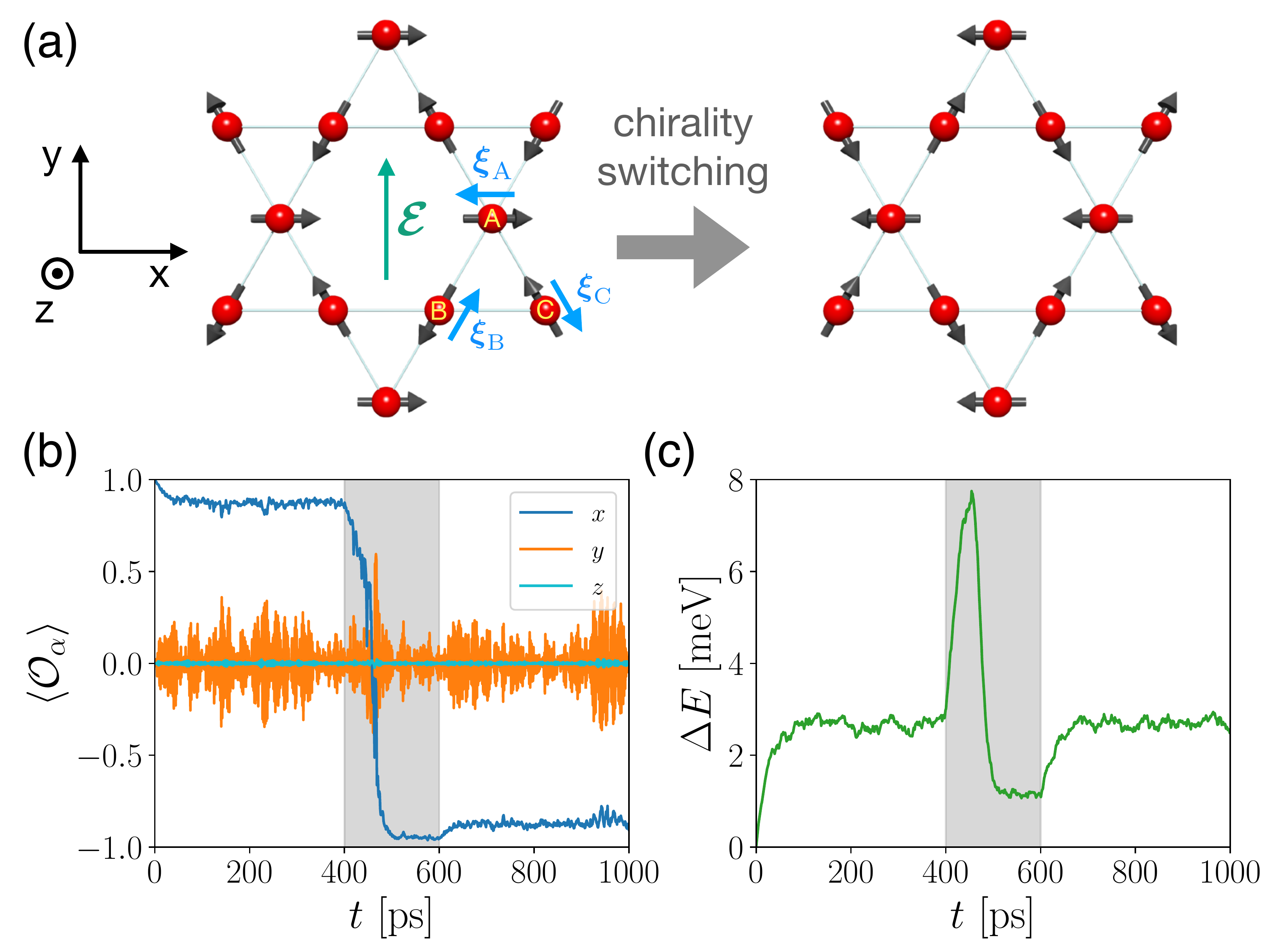}
\caption{
\label{fig:switching}
(a) Chirality switching induced by the NC spin current, where black arrows represent local moments of Mn atoms in Mn$_3$X. An external electric field $\boldsymbol{\mathcal{E}}$ is indicated by a green arrow.  Time evolution of (b) the average octupole moments and (c) energy per unit cell obtained from atomistic spin dynamics simulation. The chiral self-induced torque is applied during $t=400-600\ \mathrm{ps}$. 
}
\end{figure}

In a Mn$_3$X film grown along [0001] direction (perpendicular to the Kagome plane), the chiral component of the NC spin current induced by an external electric field leads to a spin accumulation at both surfaces, which exerts a torque on local moments. This is analogous to the self-induced torque at the interface of a single ferromagnet \cite{Wang2019, Cespedes-Berrocal2021}. While the self-induced torque cancels to zero for a stand-alone film, interaction of the Mn$_3$X film and the substrate will cause an asymmetry between the top and bottom surfaces, leading to finite self-induced torque in a ``chiral'' manner. We propose an idea that the self-induced torque caused by the NC spin current can be used to switch the chiral magnetic texture of Mn$_3$X as illustrated in Fig.~\ref{fig:switching}(a).

In order to demonstrate the switching of the chiral spin texture, we consider a classical spin model and perform the atomistic spin dynamics  simulations \cite{Mueller2019}. For simplicity, we consider a single layer Kagome plane with three spins in the unit cell. The Hamiltonian for the spin system is given by
\begin{eqnarray}
\mathcal{H}_\mathrm{mag}
=
& &
J \sum_{\langle ij \rangle}
\hat{\mathbf{m}}_i \cdot \hat{\mathbf{m}}_j
-\frac{K}{2}
\sum_i
\left( 
\hat{\mathbf{K}}_i \cdot \hat{\mathbf{m}}_i
\right)^2
\nonumber
\\
& & +
D \sum_{\langle ij \rangle} 
\hat{\mathbf{n}}_{ij}
\cdot (\hat{\mathbf{m}}_i \times \hat{\mathbf{m}}_j)
\label{eq:spin_Hamiltonian}
\end{eqnarray}
where $\hat{\mathbf{m}}_i$ is the direction of the magnetic moment at site $i$, $J>0$ is the strength of the exchange interaction (antiferromagnetic), $K>0$ is the strength of the in-plane anisotropy whose direction depends on the sublattice such that $\hat{\mathbf{K}}_\mathrm{A} = \hat{\mathbf{x}}$, $\hat{\mathbf{K}}_\mathrm{B} = -(1/2)\hat{\mathbf{x}}-(\sqrt{3}/2)\hat{\mathbf{y}}$, $\hat{\mathbf{K}}_\mathrm{C} = -(1/2)\hat{\mathbf{x}}+(\sqrt{3}/2)\hat{\mathbf{y}}$, and $D$ is the strength of the Dyzaloshinskii-Moriya interaction. The definition of $\hat{\mathbf{n}}_{ij}$ is identical to that in Eq.~\eqref{eq:Hamiltonian}. 

Starting from Eq.~\eqref{eq:spin_Hamiltonian}, the Landau-Lifshitz-Gilbert (LLG) equation can be written as:
\begin{eqnarray}
\frac{d\hat{\mathbf{m}}_i}{dt}
=
-|\gamma|\hat{\mathbf{m}}_i \times \left( \mathbf{B}_i^\mathrm{eff} + \mathbf{B}_i^\mathrm{fl} \right)+ 
\alpha
\hat{\mathbf{m}}_i \times \frac{d\hat{\mathbf{m}}_i}{dt}
+
\boldsymbol{\tau}_i^\mathrm{chiral},
\nonumber
\\
\label{eq:LLG}
\end{eqnarray}
where $|\gamma|$ is the magnitude of the gyromagnetic ratio for an electron, $\mathbf{B}_i^\mathrm{eff}$ is the effective magnetic field, $\mathbf{B}_i^\mathrm{fl}$ is a random field induced by thermal fluctuation of the system, and $\alpha$ is Gilbert damping constant. The effective field is given by $\mathbf{B}_i^\mathrm{eff} = -(1/m_0)(\delta \mathcal{H}_\mathrm{mag}/\delta \mathbf{m}_i)$, where $m_0$ is the magnitude of the magnetic moment. The fluctuating field is given as a Gaussian noise such that $\langle B_{i,\alpha}^\mathrm{fl} (t) B_{j,\beta}^\mathrm{fl} (t') \rangle = 2 \delta_{ij} \delta_{\alpha\beta} \delta (t-t') \alpha k_\mathrm{B} T/m_0|\gamma|$ for $\alpha,\beta = x,y,z$, where $k_\mathrm{B}$ is the Boltzmann constant and $T$ is the temperature. We apply the torque that is induced by the injection of the NC spin current for the chiral component only. It has a form 
$\boldsymbol{\tau}_i^\mathrm{chiral}
=
(\tau_0|\gamma|/\mu_\mathrm{B}) \hat{\mathbf{m}}_i \times  (\hat{\boldsymbol{\xi}}_i\times \hat{\mathbf{m}}_i)
$, where $\tau_0$ is the magnitude of the torque, $\mu_\mathrm{B}$ is the Bohr magneton, and $\hat{\boldsymbol{\xi}}_i$'s are the chiral directions defined in Eq.~\eqref{eq:chiral_SHC}, which are indicated by blue arrows in Fig.~\ref{fig:switching}(a). This torque enables switching of the chiral magnetic texture via a soft mode that coherently rotates the magnetic moments in the Kagome plane \cite{Dasgupta2020, Chen2020}.

For the atomistic spin simulation, we choose $J=10$, $K=0.1$, $D=0.7$ in units of $\mathrm{meV}$, $m_0=3\mu_\mathrm{B}$, $\alpha=0.001$ and $T=10 \ \mathrm{K}$. The magnitude of the torque is set $\tau_0 = 0.01\ \mathrm{meV}$. We consider a $10\times 10$ supercell with the periodic boundary condition and solve the LLG equation [Eq.~\eqref{eq:LLG}] for 300 spins. For the analysis, we define an order parameter of Mn$_3$X by
\begin{eqnarray}
\boldsymbol{\mathcal{O}} = \frac{1}{3}
\left(
\hat{\mathbf{m}}_\mathrm{A}
+
R\hat{\mathbf{m}}_\mathrm{B}
+
R^2\hat{\mathbf{m}}_\mathrm{C}
\right),
\end{eqnarray}
where $R$ is an anti-clockwise rotation by $2\pi/3$ around the $z$ axis \cite{Gomonay1992}. For the ground state configurations shown in Fig.~\ref{fig:switching}, for example, $\boldsymbol{\mathcal{O}}=\hat{\mathbf{m}}_\mathrm{A}$ when $T=0$. We remark that $\boldsymbol{\mathcal{O}}$ is often referred to as the octupole magnetic moment in the literature~\cite{Higo2018, Tsai2020, Suzuki2017, Nomoto2020}. 

By numerically solving Eq.~\eqref{eq:LLG}, we show the evolution of $\boldsymbol{\mathcal{O}}$ averaged over the supercell [Fig.~\ref{fig:switching}(b)]. The torque is applied from $t=400\ \mathrm{ps}$ till $t=600\ \mathrm{ps}$ (marked by grey color shade). A plot for $\mathcal{O}_x$ (blue line) clearly shows that the chiral magnetic texture is switched by the torque within $\sim 100\ \mathrm{ps}$ interval. Meanwhile we observe fluctuation of the other components. Overall, $\mathcal{O}_y$ exhibits stronger fluctuation than $\mathcal{O}_z$, which is attributed to the in-plane anisotropy of the system. It is interesting to notice that the fluctuation becomes enhanced during the switching ($t\approx 450 \ \mathrm{ps}$). Evaluation of the energy in each time step by Eq.~\eqref{eq:spin_Hamiltonian} is shown in Fig.~\ref{fig:switching}(c). It reveals the activation energy required for the switching is $\Delta E \approx 8\ \mathrm{meV}$ per unit cell. While the overall background fluctuation of $\Delta E \approx 2.5\ \mathrm{meV}$ is due to thermal effect, the energy fluctuations becomes suppressed when the torque is still applied after the switching at $t\approx 500\ \mathrm{ps}$. 


In conclusion, we propose a concept of a NC spin current, which can be excited by the SHE in crystals with low \emph{local} symmetry such as Mn$_3$X. As it arises from the sublattice-dependent crystal potential, we expect that the NC spin current can be found in many other materials, where the crystal potentially is locally asymmetric
such as e.g. altermagnets~\cite{Smejkal2021}. A major consequence of the NC spin current results from its coupling with the chiral magnetic texture in NC AFMs. For example, in a thin film of Mn$_3$X grown on a substrate, the NC spin current may result in a self-induced torque at the interface and switch the chirality of the magnetic texture. This opens a novel route toward electric control of NC AFMs, which has been one of the most challenging problems in antiferromagnetic spintronics.  As such our finding is an important manifestation of a sublattice-dependent electronic excitations and its coupling to a magnetic texture, which is ``hidden'' in the description of \emph{global} spin current, and it is expected to play a crucial role in understanding complex nature of spin excitations in chiral magnets.

\begin{acknowledgements}
We thank Hiroshi Katsumoto, Adithya Rajan, Arnab Bose, Tom G. Saunderson, and Mathias Kl\"aui for fruitful discussions. We gratefully acknowledge the J\"ulich Supercomputing Centre for providing computational resources under project jiff40. This work was funded by the Deutsche Forschungsgemeinschaft (DFG, German Research Foundation) $-$ TRR 173/2 $-$ 268565370 Spin+X (Project A11 and B12), TRR 288 $-$ 422213477 (Project B06), and the Sino-German research project DISTOMAT (MO 1731/10-1). O.G. acknowledges EU FET Open RIA Grant no. 766566.
\end{acknowledgements}

\bibliography{reference}

\begin{thebibliography}{41}%
\makeatletter
\providecommand \@ifxundefined [1]{%
 \@ifx{#1\undefined}
}%
\providecommand \@ifnum [1]{%
 \ifnum #1\expandafter \@firstoftwo
 \else \expandafter \@secondoftwo
 \fi
}%
\providecommand \@ifx [1]{%
 \ifx #1\expandafter \@firstoftwo
 \else \expandafter \@secondoftwo
 \fi
}%
\providecommand \natexlab [1]{#1}%
\providecommand \enquote  [1]{``#1''}%
\providecommand \bibnamefont  [1]{#1}%
\providecommand \bibfnamefont [1]{#1}%
\providecommand \citenamefont [1]{#1}%
\providecommand \href@noop [0]{\@secondoftwo}%
\providecommand \href [0]{\begingroup \@sanitize@url \@href}%
\providecommand \@href[1]{\@@startlink{#1}\@@href}%
\providecommand \@@href[1]{\endgroup#1\@@endlink}%
\providecommand \@sanitize@url [0]{\catcode `\\12\catcode `\$12\catcode
  `\&12\catcode `\#12\catcode `\^12\catcode `\_12\catcode `\%12\relax}%
\providecommand \@@startlink[1]{}%
\providecommand \@@endlink[0]{}%
\providecommand \url  [0]{\begingroup\@sanitize@url \@url }%
\providecommand \@url [1]{\endgroup\@href {#1}{\urlprefix }}%
\providecommand \urlprefix  [0]{URL }%
\providecommand \Eprint [0]{\href }%
\providecommand \doibase [0]{https://doi.org/}%
\providecommand \selectlanguage [0]{\@gobble}%
\providecommand \bibinfo  [0]{\@secondoftwo}%
\providecommand \bibfield  [0]{\@secondoftwo}%
\providecommand \translation [1]{[#1]}%
\providecommand \BibitemOpen [0]{}%
\providecommand \bibitemStop [0]{}%
\providecommand \bibitemNoStop [0]{.\EOS\space}%
\providecommand \EOS [0]{\spacefactor3000\relax}%
\providecommand \BibitemShut  [1]{\csname bibitem#1\endcsname}%
\let\auto@bib@innerbib\@empty
\bibitem [{\citenamefont {Jungwirth}\ \emph {et~al.}(2016)\citenamefont
  {Jungwirth}, \citenamefont {Marti}, \citenamefont {Wadley},\ and\
  \citenamefont {Wunderlich}}]{Jungwirth2016}%
  \BibitemOpen
  \bibfield  {author} {\bibinfo {author} {\bibfnamefont {T.}~\bibnamefont
  {Jungwirth}}, \bibinfo {author} {\bibfnamefont {X.}~\bibnamefont {Marti}},
  \bibinfo {author} {\bibfnamefont {P.}~\bibnamefont {Wadley}},\ and\ \bibinfo
  {author} {\bibfnamefont {J.}~\bibnamefont {Wunderlich}},\ }\bibfield  {title}
  {\bibinfo {title} {Antiferromagnetic spintronics},\ }\href
  {https://doi.org/10.1038/nnano.2016.18} {\bibfield  {journal} {\bibinfo
  {journal} {Nature Nanotechnology}\ }\textbf {\bibinfo {volume} {11}},\
  \bibinfo {pages} {231} (\bibinfo {year} {2016})}\BibitemShut {NoStop}%
\bibitem [{\citenamefont {Baltz}\ \emph {et~al.}(2018)\citenamefont {Baltz},
  \citenamefont {Manchon}, \citenamefont {Tsoi}, \citenamefont {Moriyama},
  \citenamefont {Ono},\ and\ \citenamefont {Tserkovnyak}}]{Baltz2018}%
  \BibitemOpen
  \bibfield  {author} {\bibinfo {author} {\bibfnamefont {V.}~\bibnamefont
  {Baltz}}, \bibinfo {author} {\bibfnamefont {A.}~\bibnamefont {Manchon}},
  \bibinfo {author} {\bibfnamefont {M.}~\bibnamefont {Tsoi}}, \bibinfo {author}
  {\bibfnamefont {T.}~\bibnamefont {Moriyama}}, \bibinfo {author}
  {\bibfnamefont {T.}~\bibnamefont {Ono}},\ and\ \bibinfo {author}
  {\bibfnamefont {Y.}~\bibnamefont {Tserkovnyak}},\ }\bibfield  {title}
  {\bibinfo {title} {Antiferromagnetic spintronics},\ }\href
  {https://doi.org/10.1103/RevModPhys.90.015005} {\bibfield  {journal}
  {\bibinfo  {journal} {Rev. Mod. Phys.}\ }\textbf {\bibinfo {volume} {90}},\
  \bibinfo {pages} {015005} (\bibinfo {year} {2018})}\BibitemShut {NoStop}%
\bibitem [{\citenamefont {Keffer}\ and\ \citenamefont
  {Kittel}(1952)}]{Kittel1952}%
  \BibitemOpen
  \bibfield  {author} {\bibinfo {author} {\bibfnamefont {F.}~\bibnamefont
  {Keffer}}\ and\ \bibinfo {author} {\bibfnamefont {C.}~\bibnamefont
  {Kittel}},\ }\bibfield  {title} {\bibinfo {title} {Theory of
  antiferromagnetic resonance},\ }\href
  {https://doi.org/10.1103/PhysRev.85.329} {\bibfield  {journal} {\bibinfo
  {journal} {Phys. Rev.}\ }\textbf {\bibinfo {volume} {85}},\ \bibinfo {pages}
  {329} (\bibinfo {year} {1952})}\BibitemShut {NoStop}%
\bibitem [{\citenamefont {Ross}\ \emph {et~al.}(2015)\citenamefont {Ross},
  \citenamefont {Schreier}, \citenamefont {Lotze}, \citenamefont {Huebl},
  \citenamefont {Gross},\ and\ \citenamefont {Goennenwein}}]{Ross2015}%
  \BibitemOpen
  \bibfield  {author} {\bibinfo {author} {\bibfnamefont {P.}~\bibnamefont
  {Ross}}, \bibinfo {author} {\bibfnamefont {M.}~\bibnamefont {Schreier}},
  \bibinfo {author} {\bibfnamefont {J.}~\bibnamefont {Lotze}}, \bibinfo
  {author} {\bibfnamefont {H.}~\bibnamefont {Huebl}}, \bibinfo {author}
  {\bibfnamefont {R.}~\bibnamefont {Gross}},\ and\ \bibinfo {author}
  {\bibfnamefont {S.~T.~B.}\ \bibnamefont {Goennenwein}},\ }\bibfield  {title}
  {\bibinfo {title} {{Antiferromagentic resonance detected by direct current
  voltages in MnF$_2$/Pt bilayers}},\ }\href
  {https://doi.org/10.1063/1.4937913} {\bibfield  {journal} {\bibinfo
  {journal} {Journal of Applied Physics}\ }\textbf {\bibinfo {volume} {118}},\
  \bibinfo {pages} {233907} (\bibinfo {year} {2015})}\BibitemShut {NoStop}%
\bibitem [{\citenamefont {Gomonay}\ \emph {et~al.}(2018)\citenamefont
  {Gomonay}, \citenamefont {Baltz}, \citenamefont {Brataas},\ and\
  \citenamefont {Tserkovnyak}}]{Gomonay2018}%
  \BibitemOpen
  \bibfield  {author} {\bibinfo {author} {\bibfnamefont {O.}~\bibnamefont
  {Gomonay}}, \bibinfo {author} {\bibfnamefont {V.}~\bibnamefont {Baltz}},
  \bibinfo {author} {\bibfnamefont {A.}~\bibnamefont {Brataas}},\ and\ \bibinfo
  {author} {\bibfnamefont {Y.}~\bibnamefont {Tserkovnyak}},\ }\bibfield
  {title} {\bibinfo {title} {Antiferromagnetic spin textures and dynamics},\
  }\href {https://doi.org/10.1038/s41567-018-0049-4} {\bibfield  {journal}
  {\bibinfo  {journal} {Nature Physics}\ }\textbf {\bibinfo {volume} {14}},\
  \bibinfo {pages} {213} (\bibinfo {year} {2018})}\BibitemShut {NoStop}%
\bibitem [{\citenamefont {\ifmmode~\check{Z}\else \v{Z}\fi{}elezn\'y}\ \emph
  {et~al.}(2014)\citenamefont {\ifmmode~\check{Z}\else \v{Z}\fi{}elezn\'y},
  \citenamefont {Gao}, \citenamefont {V\'yborn\'y}, \citenamefont {Zemen},
  \citenamefont {Ma\ifmmode~\check{s}\else \v{s}\fi{}ek}, \citenamefont
  {Manchon}, \citenamefont {Wunderlich}, \citenamefont {Sinova},\ and\
  \citenamefont {Jungwirth}}]{Zelezny2014}%
  \BibitemOpen
  \bibfield  {author} {\bibinfo {author} {\bibfnamefont {J.}~\bibnamefont
  {\ifmmode~\check{Z}\else \v{Z}\fi{}elezn\'y}}, \bibinfo {author}
  {\bibfnamefont {H.}~\bibnamefont {Gao}}, \bibinfo {author} {\bibfnamefont
  {K.}~\bibnamefont {V\'yborn\'y}}, \bibinfo {author} {\bibfnamefont
  {J.}~\bibnamefont {Zemen}}, \bibinfo {author} {\bibfnamefont
  {J.}~\bibnamefont {Ma\ifmmode~\check{s}\else \v{s}\fi{}ek}}, \bibinfo
  {author} {\bibfnamefont {A.}~\bibnamefont {Manchon}}, \bibinfo {author}
  {\bibfnamefont {J.}~\bibnamefont {Wunderlich}}, \bibinfo {author}
  {\bibfnamefont {J.}~\bibnamefont {Sinova}},\ and\ \bibinfo {author}
  {\bibfnamefont {T.}~\bibnamefont {Jungwirth}},\ }\bibfield  {title} {\bibinfo
  {title} {{Relativistic N\'eel-Order Fields Induced by Electrical Current in
  Antiferromagnets}},\ }\href {https://doi.org/10.1103/PhysRevLett.113.157201}
  {\bibfield  {journal} {\bibinfo  {journal} {Phys. Rev. Lett.}\ }\textbf
  {\bibinfo {volume} {113}},\ \bibinfo {pages} {157201} (\bibinfo {year}
  {2014})}\BibitemShut {NoStop}%
\bibitem [{\citenamefont {Wadley}\ \emph {et~al.}(2016)\citenamefont {Wadley},
  \citenamefont {Howells}, \citenamefont {Železný}, \citenamefont {Andrews},
  \citenamefont {Hills}, \citenamefont {Campion}, \citenamefont {Novák},
  \citenamefont {Olejník}, \citenamefont {Maccherozzi}, \citenamefont {Dhesi},
  \citenamefont {Martin}, \citenamefont {Wagner}, \citenamefont {Wunderlich},
  \citenamefont {Freimuth}, \citenamefont {Mokrousov}, \citenamefont {Kuneš},
  \citenamefont {Chauhan}, \citenamefont {Grzybowski}, \citenamefont
  {Rushforth}, \citenamefont {Edmonds}, \citenamefont {Gallagher},\ and\
  \citenamefont {Jungwirth}}]{Wadley2016}%
  \BibitemOpen
  \bibfield  {author} {\bibinfo {author} {\bibfnamefont {P.}~\bibnamefont
  {Wadley}}, \bibinfo {author} {\bibfnamefont {B.}~\bibnamefont {Howells}},
  \bibinfo {author} {\bibfnamefont {J.}~\bibnamefont {Železný}}, \bibinfo
  {author} {\bibfnamefont {C.}~\bibnamefont {Andrews}}, \bibinfo {author}
  {\bibfnamefont {V.}~\bibnamefont {Hills}}, \bibinfo {author} {\bibfnamefont
  {R.~P.}\ \bibnamefont {Campion}}, \bibinfo {author} {\bibfnamefont
  {V.}~\bibnamefont {Novák}}, \bibinfo {author} {\bibfnamefont
  {K.}~\bibnamefont {Olejník}}, \bibinfo {author} {\bibfnamefont
  {F.}~\bibnamefont {Maccherozzi}}, \bibinfo {author} {\bibfnamefont {S.~S.}\
  \bibnamefont {Dhesi}}, \bibinfo {author} {\bibfnamefont {S.~Y.}\ \bibnamefont
  {Martin}}, \bibinfo {author} {\bibfnamefont {T.}~\bibnamefont {Wagner}},
  \bibinfo {author} {\bibfnamefont {J.}~\bibnamefont {Wunderlich}}, \bibinfo
  {author} {\bibfnamefont {F.}~\bibnamefont {Freimuth}}, \bibinfo {author}
  {\bibfnamefont {Y.}~\bibnamefont {Mokrousov}}, \bibinfo {author}
  {\bibfnamefont {J.}~\bibnamefont {Kuneš}}, \bibinfo {author} {\bibfnamefont
  {J.~S.}\ \bibnamefont {Chauhan}}, \bibinfo {author} {\bibfnamefont {M.~J.}\
  \bibnamefont {Grzybowski}}, \bibinfo {author} {\bibfnamefont {A.~W.}\
  \bibnamefont {Rushforth}}, \bibinfo {author} {\bibfnamefont {K.~W.}\
  \bibnamefont {Edmonds}}, \bibinfo {author} {\bibfnamefont {B.~L.}\
  \bibnamefont {Gallagher}},\ and\ \bibinfo {author} {\bibfnamefont
  {T.}~\bibnamefont {Jungwirth}},\ }\bibfield  {title} {\bibinfo {title}
  {Electrical switching of an antiferromagnet},\ }\href
  {https://doi.org/10.1126/science.aab1031} {\bibfield  {journal} {\bibinfo
  {journal} {Science}\ }\textbf {\bibinfo {volume} {351}},\ \bibinfo {pages}
  {587} (\bibinfo {year} {2016})}\BibitemShut {NoStop}%
\bibitem [{\citenamefont {\ifmmode~\check{Z}\else \v{Z}\fi{}elezn\'y}\ \emph
  {et~al.}(2017{\natexlab{a}})\citenamefont {\ifmmode~\check{Z}\else
  \v{Z}\fi{}elezn\'y}, \citenamefont {Gao}, \citenamefont {Manchon},
  \citenamefont {Freimuth}, \citenamefont {Mokrousov}, \citenamefont {Zemen},
  \citenamefont {Ma\ifmmode~\check{s}\else \v{s}\fi{}ek}, \citenamefont
  {Sinova},\ and\ \citenamefont {Jungwirth}}]{Zelezny2017}%
  \BibitemOpen
  \bibfield  {author} {\bibinfo {author} {\bibfnamefont {J.}~\bibnamefont
  {\ifmmode~\check{Z}\else \v{Z}\fi{}elezn\'y}}, \bibinfo {author}
  {\bibfnamefont {H.}~\bibnamefont {Gao}}, \bibinfo {author} {\bibfnamefont
  {A.}~\bibnamefont {Manchon}}, \bibinfo {author} {\bibfnamefont
  {F.}~\bibnamefont {Freimuth}}, \bibinfo {author} {\bibfnamefont
  {Y.}~\bibnamefont {Mokrousov}}, \bibinfo {author} {\bibfnamefont
  {J.}~\bibnamefont {Zemen}}, \bibinfo {author} {\bibfnamefont
  {J.}~\bibnamefont {Ma\ifmmode~\check{s}\else \v{s}\fi{}ek}}, \bibinfo
  {author} {\bibfnamefont {J.}~\bibnamefont {Sinova}},\ and\ \bibinfo {author}
  {\bibfnamefont {T.}~\bibnamefont {Jungwirth}},\ }\bibfield  {title} {\bibinfo
  {title} {Spin-orbit torques in locally and globally noncentrosymmetric
  crystals: Antiferromagnets and ferromagnets},\ }\href
  {https://doi.org/10.1103/PhysRevB.95.014403} {\bibfield  {journal} {\bibinfo
  {journal} {Phys. Rev. B}\ }\textbf {\bibinfo {volume} {95}},\ \bibinfo
  {pages} {014403} (\bibinfo {year} {2017}{\natexlab{a}})}\BibitemShut
  {NoStop}%
\bibitem [{\citenamefont {Bodnar}\ \emph {et~al.}(2018)\citenamefont {Bodnar},
  \citenamefont {{\v{S}}mejkal}, \citenamefont {Turek}, \citenamefont
  {Jungwirth}, \citenamefont {Gomonay}, \citenamefont {Sinova}, \citenamefont
  {Sapozhnik}, \citenamefont {Elmers}, \citenamefont {Kl{\"a}ui},\ and\
  \citenamefont {Jourdan}}]{Bodnar2018}%
  \BibitemOpen
  \bibfield  {author} {\bibinfo {author} {\bibfnamefont {S.~Y.}\ \bibnamefont
  {Bodnar}}, \bibinfo {author} {\bibfnamefont {L.}~\bibnamefont
  {{\v{S}}mejkal}}, \bibinfo {author} {\bibfnamefont {I.}~\bibnamefont
  {Turek}}, \bibinfo {author} {\bibfnamefont {T.}~\bibnamefont {Jungwirth}},
  \bibinfo {author} {\bibfnamefont {O.}~\bibnamefont {Gomonay}}, \bibinfo
  {author} {\bibfnamefont {J.}~\bibnamefont {Sinova}}, \bibinfo {author}
  {\bibfnamefont {A.~A.}\ \bibnamefont {Sapozhnik}}, \bibinfo {author}
  {\bibfnamefont {H.-J.}\ \bibnamefont {Elmers}}, \bibinfo {author}
  {\bibfnamefont {M.}~\bibnamefont {Kl{\"a}ui}},\ and\ \bibinfo {author}
  {\bibfnamefont {M.}~\bibnamefont {Jourdan}},\ }\bibfield  {title} {\bibinfo
  {title} {{Writing and reading antiferromagnetic Mn$_2$Au by N{\'e}el
  spin-orbit torques and large anisotropic magnetoresistance}},\ }\href
  {https://doi.org/10.1038/s41467-017-02780-x} {\bibfield  {journal} {\bibinfo
  {journal} {Nature Communications}\ }\textbf {\bibinfo {volume} {9}},\
  \bibinfo {pages} {348} (\bibinfo {year} {2018})}\BibitemShut {NoStop}%
\bibitem [{\citenamefont {Meinert}\ \emph {et~al.}(2018)\citenamefont
  {Meinert}, \citenamefont {Graulich},\ and\ \citenamefont
  {Matalla-Wagner}}]{Meinert2018}%
  \BibitemOpen
  \bibfield  {author} {\bibinfo {author} {\bibfnamefont {M.}~\bibnamefont
  {Meinert}}, \bibinfo {author} {\bibfnamefont {D.}~\bibnamefont {Graulich}},\
  and\ \bibinfo {author} {\bibfnamefont {T.}~\bibnamefont {Matalla-Wagner}},\
  }\bibfield  {title} {\bibinfo {title} {{Electrical Switching of
  Antiferromagnetic ${\mathrm{Mn}}_{2}\mathrm{Au}$ and the Role of Thermal
  Activation}},\ }\href {https://doi.org/10.1103/PhysRevApplied.9.064040}
  {\bibfield  {journal} {\bibinfo  {journal} {Phys. Rev. Applied}\ }\textbf
  {\bibinfo {volume} {9}},\ \bibinfo {pages} {064040} (\bibinfo {year}
  {2018})}\BibitemShut {NoStop}%
\bibitem [{\citenamefont {Zhou}\ \emph {et~al.}(2018)\citenamefont {Zhou},
  \citenamefont {Zhang}, \citenamefont {Li}, \citenamefont {Chen},
  \citenamefont {Shi}, \citenamefont {Tan}, \citenamefont {Gu}, \citenamefont
  {Saleem}, \citenamefont {Wu}, \citenamefont {Pan},\ and\ \citenamefont
  {Song}}]{Zhou2018}%
  \BibitemOpen
  \bibfield  {author} {\bibinfo {author} {\bibfnamefont {X.~F.}\ \bibnamefont
  {Zhou}}, \bibinfo {author} {\bibfnamefont {J.}~\bibnamefont {Zhang}},
  \bibinfo {author} {\bibfnamefont {F.}~\bibnamefont {Li}}, \bibinfo {author}
  {\bibfnamefont {X.~Z.}\ \bibnamefont {Chen}}, \bibinfo {author}
  {\bibfnamefont {G.~Y.}\ \bibnamefont {Shi}}, \bibinfo {author} {\bibfnamefont
  {Y.~Z.}\ \bibnamefont {Tan}}, \bibinfo {author} {\bibfnamefont {Y.~D.}\
  \bibnamefont {Gu}}, \bibinfo {author} {\bibfnamefont {M.~S.}\ \bibnamefont
  {Saleem}}, \bibinfo {author} {\bibfnamefont {H.~Q.}\ \bibnamefont {Wu}},
  \bibinfo {author} {\bibfnamefont {F.}~\bibnamefont {Pan}},\ and\ \bibinfo
  {author} {\bibfnamefont {C.}~\bibnamefont {Song}},\ }\bibfield  {title}
  {\bibinfo {title} {{Strong Orientation-Dependent Spin-Orbit Torque in Thin
  Films of the Antiferromagnet ${\mathrm{Mn}}_{2}\mathrm{Au}$}},\ }\href
  {https://doi.org/10.1103/PhysRevApplied.9.054028} {\bibfield  {journal}
  {\bibinfo  {journal} {Phys. Rev. Applied}\ }\textbf {\bibinfo {volume} {9}},\
  \bibinfo {pages} {054028} (\bibinfo {year} {2018})}\BibitemShut {NoStop}%
\bibitem [{\citenamefont {Chen}\ \emph {et~al.}(2014)\citenamefont {Chen},
  \citenamefont {Niu},\ and\ \citenamefont {MacDonald}}]{Chen2014}%
  \BibitemOpen
  \bibfield  {author} {\bibinfo {author} {\bibfnamefont {H.}~\bibnamefont
  {Chen}}, \bibinfo {author} {\bibfnamefont {Q.}~\bibnamefont {Niu}},\ and\
  \bibinfo {author} {\bibfnamefont {A.~H.}\ \bibnamefont {MacDonald}},\
  }\bibfield  {title} {\bibinfo {title} {{Anomalous Hall Effect Arising from
  Noncollinear Antiferromagnetism}},\ }\href
  {https://doi.org/10.1103/PhysRevLett.112.017205} {\bibfield  {journal}
  {\bibinfo  {journal} {Phys. Rev. Lett.}\ }\textbf {\bibinfo {volume} {112}},\
  \bibinfo {pages} {017205} (\bibinfo {year} {2014})}\BibitemShut {NoStop}%
\bibitem [{\citenamefont {Kübler}\ and\ \citenamefont
  {Felser}(2014)}]{Kubler2014}%
  \BibitemOpen
  \bibfield  {author} {\bibinfo {author} {\bibfnamefont {J.}~\bibnamefont
  {Kübler}}\ and\ \bibinfo {author} {\bibfnamefont {C.}~\bibnamefont
  {Felser}},\ }\bibfield  {title} {\bibinfo {title} {{Non-collinear
  antiferromagnets and the anomalous Hall effect}},\ }\href
  {https://doi.org/10.1209/0295-5075/108/67001} {\bibfield  {journal} {\bibinfo
   {journal} {{EPL} (Europhysics Letters)}\ }\textbf {\bibinfo {volume}
  {108}},\ \bibinfo {pages} {67001} (\bibinfo {year} {2014})}\BibitemShut
  {NoStop}%
\bibitem [{\citenamefont {Nakatsuji}\ \emph {et~al.}(2015)\citenamefont
  {Nakatsuji}, \citenamefont {Kiyohara},\ and\ \citenamefont
  {Higo}}]{Nakatsuji2015}%
  \BibitemOpen
  \bibfield  {author} {\bibinfo {author} {\bibfnamefont {S.}~\bibnamefont
  {Nakatsuji}}, \bibinfo {author} {\bibfnamefont {N.}~\bibnamefont
  {Kiyohara}},\ and\ \bibinfo {author} {\bibfnamefont {T.}~\bibnamefont
  {Higo}},\ }\bibfield  {title} {\bibinfo {title} {{Large anomalous Hall effect
  in a non-collinear antiferromagnet at room temperature}},\ }\href
  {https://doi.org/10.1038/nature15723} {\bibfield  {journal} {\bibinfo
  {journal} {Nature}\ }\textbf {\bibinfo {volume} {527}},\ \bibinfo {pages}
  {212} (\bibinfo {year} {2015})}\BibitemShut {NoStop}%
\bibitem [{\citenamefont {Nayak}\ \emph {et~al.}(2016)\citenamefont {Nayak},
  \citenamefont {Fischer}, \citenamefont {Sun}, \citenamefont {Yan},
  \citenamefont {Karel}, \citenamefont {Komarek}, \citenamefont {Shekhar},
  \citenamefont {Kumar}, \citenamefont {Schnelle}, \citenamefont {Kübler},
  \citenamefont {Felser},\ and\ \citenamefont {Parkin}}]{Nayak2016}%
  \BibitemOpen
  \bibfield  {author} {\bibinfo {author} {\bibfnamefont {A.~K.}\ \bibnamefont
  {Nayak}}, \bibinfo {author} {\bibfnamefont {J.~E.}\ \bibnamefont {Fischer}},
  \bibinfo {author} {\bibfnamefont {Y.}~\bibnamefont {Sun}}, \bibinfo {author}
  {\bibfnamefont {B.}~\bibnamefont {Yan}}, \bibinfo {author} {\bibfnamefont
  {J.}~\bibnamefont {Karel}}, \bibinfo {author} {\bibfnamefont {A.~C.}\
  \bibnamefont {Komarek}}, \bibinfo {author} {\bibfnamefont {C.}~\bibnamefont
  {Shekhar}}, \bibinfo {author} {\bibfnamefont {N.}~\bibnamefont {Kumar}},
  \bibinfo {author} {\bibfnamefont {W.}~\bibnamefont {Schnelle}}, \bibinfo
  {author} {\bibfnamefont {J.}~\bibnamefont {Kübler}}, \bibinfo {author}
  {\bibfnamefont {C.}~\bibnamefont {Felser}},\ and\ \bibinfo {author}
  {\bibfnamefont {S.~S.~P.}\ \bibnamefont {Parkin}},\ }\bibfield  {title}
  {\bibinfo {title} {{Large anomalous Hall effect driven by a nonvanishing
  Berry curvature in the noncolinear antiferromagnet Mn$_3$Ge}},\ }\href
  {https://doi.org/10.1126/sciadv.1501870} {\bibfield  {journal} {\bibinfo
  {journal} {Science Advances}\ }\textbf {\bibinfo {volume} {2}},\ \bibinfo
  {pages} {e1501870} (\bibinfo {year} {2016})}\BibitemShut {NoStop}%
\bibitem [{\citenamefont {Ikhlas}\ \emph {et~al.}(2017)\citenamefont {Ikhlas},
  \citenamefont {Tomita}, \citenamefont {Koretsune}, \citenamefont {Suzuki},
  \citenamefont {Nishio-Hamane}, \citenamefont {Arita}, \citenamefont {Otani},\
  and\ \citenamefont {Nakatsuji}}]{Ikhlas2017}%
  \BibitemOpen
  \bibfield  {author} {\bibinfo {author} {\bibfnamefont {M.}~\bibnamefont
  {Ikhlas}}, \bibinfo {author} {\bibfnamefont {T.}~\bibnamefont {Tomita}},
  \bibinfo {author} {\bibfnamefont {T.}~\bibnamefont {Koretsune}}, \bibinfo
  {author} {\bibfnamefont {M.-T.}\ \bibnamefont {Suzuki}}, \bibinfo {author}
  {\bibfnamefont {D.}~\bibnamefont {Nishio-Hamane}}, \bibinfo {author}
  {\bibfnamefont {R.}~\bibnamefont {Arita}}, \bibinfo {author} {\bibfnamefont
  {Y.}~\bibnamefont {Otani}},\ and\ \bibinfo {author} {\bibfnamefont
  {S.}~\bibnamefont {Nakatsuji}},\ }\bibfield  {title} {\bibinfo {title}
  {{Large anomalous Nernst effect at room temperature in a chiral
  antiferromagnet}},\ }\href {https://doi.org/10.1038/nphys4181} {\bibfield
  {journal} {\bibinfo  {journal} {Nature Physics}\ }\textbf {\bibinfo {volume}
  {13}},\ \bibinfo {pages} {1085} (\bibinfo {year} {2017})}\BibitemShut
  {NoStop}%
\bibitem [{\citenamefont {Reichlova}\ \emph {et~al.}(2019)\citenamefont
  {Reichlova}, \citenamefont {Janda}, \citenamefont {Godinho}, \citenamefont
  {Markou}, \citenamefont {Kriegner}, \citenamefont {Schlitz}, \citenamefont
  {Zelezny}, \citenamefont {Soban}, \citenamefont {Bejarano}, \citenamefont
  {Schultheiss}, \citenamefont {Nemec}, \citenamefont {Jungwirth},
  \citenamefont {Felser}, \citenamefont {Wunderlich},\ and\ \citenamefont
  {Goennenwein}}]{Reichlova2019}%
  \BibitemOpen
  \bibfield  {author} {\bibinfo {author} {\bibfnamefont {H.}~\bibnamefont
  {Reichlova}}, \bibinfo {author} {\bibfnamefont {T.}~\bibnamefont {Janda}},
  \bibinfo {author} {\bibfnamefont {J.}~\bibnamefont {Godinho}}, \bibinfo
  {author} {\bibfnamefont {A.}~\bibnamefont {Markou}}, \bibinfo {author}
  {\bibfnamefont {D.}~\bibnamefont {Kriegner}}, \bibinfo {author}
  {\bibfnamefont {R.}~\bibnamefont {Schlitz}}, \bibinfo {author} {\bibfnamefont
  {J.}~\bibnamefont {Zelezny}}, \bibinfo {author} {\bibfnamefont
  {Z.}~\bibnamefont {Soban}}, \bibinfo {author} {\bibfnamefont
  {M.}~\bibnamefont {Bejarano}}, \bibinfo {author} {\bibfnamefont
  {H.}~\bibnamefont {Schultheiss}}, \bibinfo {author} {\bibfnamefont
  {P.}~\bibnamefont {Nemec}}, \bibinfo {author} {\bibfnamefont
  {T.}~\bibnamefont {Jungwirth}}, \bibinfo {author} {\bibfnamefont
  {C.}~\bibnamefont {Felser}}, \bibinfo {author} {\bibfnamefont
  {J.}~\bibnamefont {Wunderlich}},\ and\ \bibinfo {author} {\bibfnamefont
  {S.~T.~B.}\ \bibnamefont {Goennenwein}},\ }\bibfield  {title} {\bibinfo
  {title} {{Imaging and writing magnetic domains in the non-collinear
  antiferromagnet Mn$_3$Sn}},\ }\href
  {https://doi.org/10.1038/s41467-019-13391-z} {\bibfield  {journal} {\bibinfo
  {journal} {Nature Communications}\ }\textbf {\bibinfo {volume} {10}},\
  \bibinfo {pages} {5459} (\bibinfo {year} {2019})}\BibitemShut {NoStop}%
\bibitem [{\citenamefont {Higo}\ \emph {et~al.}(2018)\citenamefont {Higo},
  \citenamefont {Man}, \citenamefont {Gopman}, \citenamefont {Wu},
  \citenamefont {Koretsune}, \citenamefont {van~'t Erve}, \citenamefont
  {Kabanov}, \citenamefont {Rees}, \citenamefont {Li}, \citenamefont {Suzuki},
  \citenamefont {Patankar}, \citenamefont {Ikhlas}, \citenamefont {Chien},
  \citenamefont {Arita}, \citenamefont {Shull}, \citenamefont {Orenstein},\
  and\ \citenamefont {Nakatsuji}}]{Higo2018}%
  \BibitemOpen
  \bibfield  {author} {\bibinfo {author} {\bibfnamefont {T.}~\bibnamefont
  {Higo}}, \bibinfo {author} {\bibfnamefont {H.}~\bibnamefont {Man}}, \bibinfo
  {author} {\bibfnamefont {D.~B.}\ \bibnamefont {Gopman}}, \bibinfo {author}
  {\bibfnamefont {L.}~\bibnamefont {Wu}}, \bibinfo {author} {\bibfnamefont
  {T.}~\bibnamefont {Koretsune}}, \bibinfo {author} {\bibfnamefont {O.~M.~J.}\
  \bibnamefont {van~'t Erve}}, \bibinfo {author} {\bibfnamefont {Y.~P.}\
  \bibnamefont {Kabanov}}, \bibinfo {author} {\bibfnamefont {D.}~\bibnamefont
  {Rees}}, \bibinfo {author} {\bibfnamefont {Y.}~\bibnamefont {Li}}, \bibinfo
  {author} {\bibfnamefont {M.-T.}\ \bibnamefont {Suzuki}}, \bibinfo {author}
  {\bibfnamefont {S.}~\bibnamefont {Patankar}}, \bibinfo {author}
  {\bibfnamefont {M.}~\bibnamefont {Ikhlas}}, \bibinfo {author} {\bibfnamefont
  {C.~L.}\ \bibnamefont {Chien}}, \bibinfo {author} {\bibfnamefont
  {R.}~\bibnamefont {Arita}}, \bibinfo {author} {\bibfnamefont {R.~D.}\
  \bibnamefont {Shull}}, \bibinfo {author} {\bibfnamefont {J.}~\bibnamefont
  {Orenstein}},\ and\ \bibinfo {author} {\bibfnamefont {S.}~\bibnamefont
  {Nakatsuji}},\ }\bibfield  {title} {\bibinfo {title} {{Large magneto-optical
  Kerr effect and imaging of magnetic octupole domains in an antiferromagnetic
  metal}},\ }\href {https://doi.org/10.1038/s41566-017-0086-z} {\bibfield
  {journal} {\bibinfo  {journal} {Nature Photonics}\ }\textbf {\bibinfo
  {volume} {12}},\ \bibinfo {pages} {73} (\bibinfo {year} {2018})}\BibitemShut
  {NoStop}%
\bibitem [{\citenamefont {Kuroda}\ \emph {et~al.}(2017)\citenamefont {Kuroda},
  \citenamefont {Tomita}, \citenamefont {Suzuki}, \citenamefont {Bareille},
  \citenamefont {Nugroho}, \citenamefont {Goswami}, \citenamefont {Ochi},
  \citenamefont {Ikhlas}, \citenamefont {Nakayama}, \citenamefont {Akebi},
  \citenamefont {Noguchi}, \citenamefont {Ishii}, \citenamefont {Inami},
  \citenamefont {Ono}, \citenamefont {Kumigashira}, \citenamefont {Varykhalov},
  \citenamefont {Muro}, \citenamefont {Koretsune}, \citenamefont {Arita},
  \citenamefont {Shin}, \citenamefont {Kondo},\ and\ \citenamefont
  {Nakatsuji}}]{Kuroda2017}%
  \BibitemOpen
  \bibfield  {author} {\bibinfo {author} {\bibfnamefont {K.}~\bibnamefont
  {Kuroda}}, \bibinfo {author} {\bibfnamefont {T.}~\bibnamefont {Tomita}},
  \bibinfo {author} {\bibfnamefont {M.-T.}\ \bibnamefont {Suzuki}}, \bibinfo
  {author} {\bibfnamefont {C.}~\bibnamefont {Bareille}}, \bibinfo {author}
  {\bibfnamefont {A.~A.}\ \bibnamefont {Nugroho}}, \bibinfo {author}
  {\bibfnamefont {P.}~\bibnamefont {Goswami}}, \bibinfo {author} {\bibfnamefont
  {M.}~\bibnamefont {Ochi}}, \bibinfo {author} {\bibfnamefont {M.}~\bibnamefont
  {Ikhlas}}, \bibinfo {author} {\bibfnamefont {M.}~\bibnamefont {Nakayama}},
  \bibinfo {author} {\bibfnamefont {S.}~\bibnamefont {Akebi}}, \bibinfo
  {author} {\bibfnamefont {R.}~\bibnamefont {Noguchi}}, \bibinfo {author}
  {\bibfnamefont {R.}~\bibnamefont {Ishii}}, \bibinfo {author} {\bibfnamefont
  {N.}~\bibnamefont {Inami}}, \bibinfo {author} {\bibfnamefont
  {K.}~\bibnamefont {Ono}}, \bibinfo {author} {\bibfnamefont {H.}~\bibnamefont
  {Kumigashira}}, \bibinfo {author} {\bibfnamefont {A.}~\bibnamefont
  {Varykhalov}}, \bibinfo {author} {\bibfnamefont {T.}~\bibnamefont {Muro}},
  \bibinfo {author} {\bibfnamefont {T.}~\bibnamefont {Koretsune}}, \bibinfo
  {author} {\bibfnamefont {R.}~\bibnamefont {Arita}}, \bibinfo {author}
  {\bibfnamefont {S.}~\bibnamefont {Shin}}, \bibinfo {author} {\bibfnamefont
  {T.}~\bibnamefont {Kondo}},\ and\ \bibinfo {author} {\bibfnamefont
  {S.}~\bibnamefont {Nakatsuji}},\ }\bibfield  {title} {\bibinfo {title}
  {{Evidence for magnetic Weyl fermions in a correlated metal}},\ }\href
  {https://doi.org/10.1038/nmat4987} {\bibfield  {journal} {\bibinfo  {journal}
  {Nature Materials}\ }\textbf {\bibinfo {volume} {16}},\ \bibinfo {pages}
  {1090} (\bibinfo {year} {2017})}\BibitemShut {NoStop}%
\bibitem [{\citenamefont {Nielsen}\ and\ \citenamefont
  {Ninomiya}(1983)}]{Nielsen1983}%
  \BibitemOpen
  \bibfield  {author} {\bibinfo {author} {\bibfnamefont {H.}~\bibnamefont
  {Nielsen}}\ and\ \bibinfo {author} {\bibfnamefont {M.}~\bibnamefont
  {Ninomiya}},\ }\bibfield  {title} {\bibinfo {title} {{The Adler-Bell-Jackiw
  anomaly and Weyl fermions in a crystal}},\ }\href
  {https://doi.org/https://doi.org/10.1016/0370-2693(83)91529-0} {\bibfield
  {journal} {\bibinfo  {journal} {Physics Letters B}\ }\textbf {\bibinfo
  {volume} {130}},\ \bibinfo {pages} {389} (\bibinfo {year}
  {1983})}\BibitemShut {NoStop}%
\bibitem [{\citenamefont {Armitage}\ \emph {et~al.}(2018)\citenamefont
  {Armitage}, \citenamefont {Mele},\ and\ \citenamefont
  {Vishwanath}}]{Armitage2018}%
  \BibitemOpen
  \bibfield  {author} {\bibinfo {author} {\bibfnamefont {N.~P.}\ \bibnamefont
  {Armitage}}, \bibinfo {author} {\bibfnamefont {E.~J.}\ \bibnamefont {Mele}},\
  and\ \bibinfo {author} {\bibfnamefont {A.}~\bibnamefont {Vishwanath}},\
  }\bibfield  {title} {\bibinfo {title} {{Weyl and Dirac semimetals in
  three-dimensional solids}},\ }\href
  {https://doi.org/10.1103/RevModPhys.90.015001} {\bibfield  {journal}
  {\bibinfo  {journal} {Rev. Mod. Phys.}\ }\textbf {\bibinfo {volume} {90}},\
  \bibinfo {pages} {015001} (\bibinfo {year} {2018})}\BibitemShut {NoStop}%
\bibitem [{\citenamefont {Tsai}\ \emph {et~al.}(2020)\citenamefont {Tsai},
  \citenamefont {Higo}, \citenamefont {Kondou}, \citenamefont {Nomoto},
  \citenamefont {Sakai}, \citenamefont {Kobayashi}, \citenamefont {Nakano},
  \citenamefont {Yakushiji}, \citenamefont {Arita}, \citenamefont {Miwa},
  \citenamefont {Otani},\ and\ \citenamefont {Nakatsuji}}]{Tsai2020}%
  \BibitemOpen
  \bibfield  {author} {\bibinfo {author} {\bibfnamefont {H.}~\bibnamefont
  {Tsai}}, \bibinfo {author} {\bibfnamefont {T.}~\bibnamefont {Higo}}, \bibinfo
  {author} {\bibfnamefont {K.}~\bibnamefont {Kondou}}, \bibinfo {author}
  {\bibfnamefont {T.}~\bibnamefont {Nomoto}}, \bibinfo {author} {\bibfnamefont
  {A.}~\bibnamefont {Sakai}}, \bibinfo {author} {\bibfnamefont
  {A.}~\bibnamefont {Kobayashi}}, \bibinfo {author} {\bibfnamefont
  {T.}~\bibnamefont {Nakano}}, \bibinfo {author} {\bibfnamefont
  {K.}~\bibnamefont {Yakushiji}}, \bibinfo {author} {\bibfnamefont
  {R.}~\bibnamefont {Arita}}, \bibinfo {author} {\bibfnamefont
  {S.}~\bibnamefont {Miwa}}, \bibinfo {author} {\bibfnamefont {Y.}~\bibnamefont
  {Otani}},\ and\ \bibinfo {author} {\bibfnamefont {S.}~\bibnamefont
  {Nakatsuji}},\ }\bibfield  {title} {\bibinfo {title} {Electrical manipulation
  of a topological antiferromagnetic state},\ }\href
  {https://doi.org/10.1038/s41586-020-2211-2} {\bibfield  {journal} {\bibinfo
  {journal} {Nature}\ }\textbf {\bibinfo {volume} {580}},\ \bibinfo {pages}
  {608} (\bibinfo {year} {2020})}\BibitemShut {NoStop}%
\bibitem [{\citenamefont {Takeuchi}\ \emph {et~al.}(2021)\citenamefont
  {Takeuchi}, \citenamefont {Yamane}, \citenamefont {Yoon}, \citenamefont
  {Itoh}, \citenamefont {Jinnai}, \citenamefont {Kanai}, \citenamefont {Ieda},
  \citenamefont {Fukami},\ and\ \citenamefont {Ohno}}]{Takeuchi2021}%
  \BibitemOpen
  \bibfield  {author} {\bibinfo {author} {\bibfnamefont {Y.}~\bibnamefont
  {Takeuchi}}, \bibinfo {author} {\bibfnamefont {Y.}~\bibnamefont {Yamane}},
  \bibinfo {author} {\bibfnamefont {J.-Y.}\ \bibnamefont {Yoon}}, \bibinfo
  {author} {\bibfnamefont {R.}~\bibnamefont {Itoh}}, \bibinfo {author}
  {\bibfnamefont {B.}~\bibnamefont {Jinnai}}, \bibinfo {author} {\bibfnamefont
  {S.}~\bibnamefont {Kanai}}, \bibinfo {author} {\bibfnamefont
  {J.}~\bibnamefont {Ieda}}, \bibinfo {author} {\bibfnamefont {S.}~\bibnamefont
  {Fukami}},\ and\ \bibinfo {author} {\bibfnamefont {H.}~\bibnamefont {Ohno}},\
  }\bibfield  {title} {\bibinfo {title} {Chiral-spin rotation of non-collinear
  antiferromagnet by spin--orbit torque},\ }\href
  {https://doi.org/10.1038/s41563-021-01005-3} {\bibfield  {journal} {\bibinfo
  {journal} {Nature Materials}\ }\textbf {\bibinfo {volume} {20}},\ \bibinfo
  {pages} {1364} (\bibinfo {year} {2021})}\BibitemShut {NoStop}%
\bibitem [{\citenamefont {Gomonay}\ and\ \citenamefont
  {Loktev}(2015)}]{Gomonay2015}%
  \BibitemOpen
  \bibfield  {author} {\bibinfo {author} {\bibfnamefont {O.~V.}\ \bibnamefont
  {Gomonay}}\ and\ \bibinfo {author} {\bibfnamefont {V.~M.}\ \bibnamefont
  {Loktev}},\ }\bibfield  {title} {\bibinfo {title} {{Using generalized
  Landau-Lifshitz equations to describe the dynamics of multi-sublattice
  antiferromagnets induced by spin-polarized current}},\ }\href
  {https://doi.org/10.1063/1.4931648} {\bibfield  {journal} {\bibinfo
  {journal} {Low Temperature Physics}\ }\textbf {\bibinfo {volume} {41}},\
  \bibinfo {pages} {698} (\bibinfo {year} {2015})}\BibitemShut {NoStop}%
\bibitem [{\citenamefont {Dasgupta}\ and\ \citenamefont
  {Tretiakov}(2022)}]{Dasgupta2022}%
  \BibitemOpen
  \bibfield  {author} {\bibinfo {author} {\bibfnamefont {S.}~\bibnamefont
  {Dasgupta}}\ and\ \bibinfo {author} {\bibfnamefont {O.~A.}\ \bibnamefont
  {Tretiakov}},\ }\href@noop {} {\bibinfo {title} {{Tuning the Hall response of
  a non-collinear antiferromagnet with spin-transfer torques}}} (\bibinfo
  {year} {2022}),\ \Eprint {https://arxiv.org/abs/arXiv:2202.06882}
  {arXiv:2202.06882} \BibitemShut {NoStop}%
\bibitem [{\citenamefont {Freimuth}\ \emph {et~al.}(2010)\citenamefont
  {Freimuth}, \citenamefont {Bl\"ugel},\ and\ \citenamefont
  {Mokrousov}}]{Freimuth2010}%
  \BibitemOpen
  \bibfield  {author} {\bibinfo {author} {\bibfnamefont {F.}~\bibnamefont
  {Freimuth}}, \bibinfo {author} {\bibfnamefont {S.}~\bibnamefont {Bl\"ugel}},\
  and\ \bibinfo {author} {\bibfnamefont {Y.}~\bibnamefont {Mokrousov}},\
  }\bibfield  {title} {\bibinfo {title} {{Anisotropic Spin Hall Effect from
  First Principles}},\ }\href {https://doi.org/10.1103/PhysRevLett.105.246602}
  {\bibfield  {journal} {\bibinfo  {journal} {Phys. Rev. Lett.}\ }\textbf
  {\bibinfo {volume} {105}},\ \bibinfo {pages} {246602} (\bibinfo {year}
  {2010})}\BibitemShut {NoStop}%
\bibitem [{\citenamefont {Rauch}\ \emph {et~al.}(2020)\citenamefont {Rauch},
  \citenamefont {T\"opler},\ and\ \citenamefont {Mertig}}]{Rauch2020}%
  \BibitemOpen
  \bibfield  {author} {\bibinfo {author} {\bibfnamefont {T.~c.~v.}\
  \bibnamefont {Rauch}}, \bibinfo {author} {\bibfnamefont {F.}~\bibnamefont
  {T\"opler}},\ and\ \bibinfo {author} {\bibfnamefont {I.}~\bibnamefont
  {Mertig}},\ }\bibfield  {title} {\bibinfo {title} {{Local spin Hall
  conductivity}},\ }\href {https://doi.org/10.1103/PhysRevB.101.064206}
  {\bibfield  {journal} {\bibinfo  {journal} {Phys. Rev. B}\ }\textbf {\bibinfo
  {volume} {101}},\ \bibinfo {pages} {064206} (\bibinfo {year}
  {2020})}\BibitemShut {NoStop}%
\bibitem [{\citenamefont {Roy}\ \emph {et~al.}(2021)\citenamefont {Roy},
  \citenamefont {Guimarães},\ and\ \citenamefont {Sławińska}}]{Roy2021}%
  \BibitemOpen
  \bibfield  {author} {\bibinfo {author} {\bibfnamefont {A.}~\bibnamefont
  {Roy}}, \bibinfo {author} {\bibfnamefont {M.~H.~D.}\ \bibnamefont
  {Guimarães}},\ and\ \bibinfo {author} {\bibfnamefont {J.}~\bibnamefont
  {Sławińska}},\ }\href@noop {} {\bibinfo {title} {{Unconventional spin Hall
  effects in nonmagnetic solids}}} (\bibinfo {year} {2021}),\ \Eprint
  {https://arxiv.org/abs/arXiv:2110.09242} {arXiv:2110.09242} \BibitemShut
  {NoStop}%
\bibitem [{\citenamefont {\ifmmode~\check{Z}\else \v{Z}\fi{}elezn\'y}\ \emph
  {et~al.}(2017{\natexlab{b}})\citenamefont {\ifmmode~\check{Z}\else
  \v{Z}\fi{}elezn\'y}, \citenamefont {Zhang}, \citenamefont {Felser},\ and\
  \citenamefont {Yan}}]{Zelezny2017b}%
  \BibitemOpen
  \bibfield  {author} {\bibinfo {author} {\bibfnamefont {J.}~\bibnamefont
  {\ifmmode~\check{Z}\else \v{Z}\fi{}elezn\'y}}, \bibinfo {author}
  {\bibfnamefont {Y.}~\bibnamefont {Zhang}}, \bibinfo {author} {\bibfnamefont
  {C.}~\bibnamefont {Felser}},\ and\ \bibinfo {author} {\bibfnamefont
  {B.}~\bibnamefont {Yan}},\ }\bibfield  {title} {\bibinfo {title}
  {{Spin-Polarized Current in Noncollinear Antiferromagnets}},\ }\href
  {https://doi.org/10.1103/PhysRevLett.119.187204} {\bibfield  {journal}
  {\bibinfo  {journal} {Phys. Rev. Lett.}\ }\textbf {\bibinfo {volume} {119}},\
  \bibinfo {pages} {187204} (\bibinfo {year} {2017}{\natexlab{b}})}\BibitemShut
  {NoStop}%
\bibitem [{\citenamefont {Kimata}\ \emph {et~al.}(2019)\citenamefont {Kimata},
  \citenamefont {Chen}, \citenamefont {Kondou}, \citenamefont {Sugimoto},
  \citenamefont {Muduli}, \citenamefont {Ikhlas}, \citenamefont {Omori},
  \citenamefont {Tomita}, \citenamefont {MacDonald}, \citenamefont
  {Nakatsuji},\ and\ \citenamefont {Otani}}]{Kimata2019}%
  \BibitemOpen
  \bibfield  {author} {\bibinfo {author} {\bibfnamefont {M.}~\bibnamefont
  {Kimata}}, \bibinfo {author} {\bibfnamefont {H.}~\bibnamefont {Chen}},
  \bibinfo {author} {\bibfnamefont {K.}~\bibnamefont {Kondou}}, \bibinfo
  {author} {\bibfnamefont {S.}~\bibnamefont {Sugimoto}}, \bibinfo {author}
  {\bibfnamefont {P.~K.}\ \bibnamefont {Muduli}}, \bibinfo {author}
  {\bibfnamefont {M.}~\bibnamefont {Ikhlas}}, \bibinfo {author} {\bibfnamefont
  {Y.}~\bibnamefont {Omori}}, \bibinfo {author} {\bibfnamefont
  {T.}~\bibnamefont {Tomita}}, \bibinfo {author} {\bibfnamefont {A.~H.}\
  \bibnamefont {MacDonald}}, \bibinfo {author} {\bibfnamefont {S.}~\bibnamefont
  {Nakatsuji}},\ and\ \bibinfo {author} {\bibfnamefont {Y.}~\bibnamefont
  {Otani}},\ }\bibfield  {title} {\bibinfo {title} {{Magnetic and magnetic
  inverse spin Hall effects in a non-collinear antiferromagnet}},\ }\href
  {https://doi.org/10.1038/s41586-018-0853-0} {\bibfield  {journal} {\bibinfo
  {journal} {Nature}\ }\textbf {\bibinfo {volume} {565}},\ \bibinfo {pages}
  {627} (\bibinfo {year} {2019})}\BibitemShut {NoStop}%
\bibitem [{\citenamefont {Mook}\ \emph {et~al.}(2020)\citenamefont {Mook},
  \citenamefont {Neumann}, \citenamefont {Johansson}, \citenamefont {Henk},\
  and\ \citenamefont {Mertig}}]{Mook2020}%
  \BibitemOpen
  \bibfield  {author} {\bibinfo {author} {\bibfnamefont {A.}~\bibnamefont
  {Mook}}, \bibinfo {author} {\bibfnamefont {R.~R.}\ \bibnamefont {Neumann}},
  \bibinfo {author} {\bibfnamefont {A.}~\bibnamefont {Johansson}}, \bibinfo
  {author} {\bibfnamefont {J.}~\bibnamefont {Henk}},\ and\ \bibinfo {author}
  {\bibfnamefont {I.}~\bibnamefont {Mertig}},\ }\bibfield  {title} {\bibinfo
  {title} {{Origin of the magnetic spin Hall effect: Spin current vorticity in
  the Fermi sea}},\ }\href {https://doi.org/10.1103/PhysRevResearch.2.023065}
  {\bibfield  {journal} {\bibinfo  {journal} {Phys. Rev. Research}\ }\textbf
  {\bibinfo {volume} {2}},\ \bibinfo {pages} {023065} (\bibinfo {year}
  {2020})}\BibitemShut {NoStop}%
\bibitem [{\citenamefont {Ito}\ and\ \citenamefont {Nomura}(2017)}]{Ito2017}%
  \BibitemOpen
  \bibfield  {author} {\bibinfo {author} {\bibfnamefont {N.}~\bibnamefont
  {Ito}}\ and\ \bibinfo {author} {\bibfnamefont {K.}~\bibnamefont {Nomura}},\
  }\bibfield  {title} {\bibinfo {title} {{Anomalous Hall Effect and Spontaneous
  Orbital Magnetization in Antiferromagnetic Weyl Metal}},\ }\href
  {https://doi.org/10.7566/JPSJ.86.063703} {\bibfield  {journal} {\bibinfo
  {journal} {Journal of the Physical Society of Japan}\ }\textbf {\bibinfo
  {volume} {86}},\ \bibinfo {pages} {063703} (\bibinfo {year}
  {2017})}\BibitemShut {NoStop}%
\bibitem [{\citenamefont {Wang}\ \emph {et~al.}(2019)\citenamefont {Wang},
  \citenamefont {Wang}, \citenamefont {Amin}, \citenamefont {Wang},
  \citenamefont {Radhakrishnan}, \citenamefont {Davidson}, \citenamefont
  {Allen}, \citenamefont {Silva}, \citenamefont {Ohldag}, \citenamefont
  {Balzar}, \citenamefont {Zink}, \citenamefont {Haney}, \citenamefont {Xiao},
  \citenamefont {Cahill}, \citenamefont {Lorenz},\ and\ \citenamefont
  {Fan}}]{Wang2019}%
  \BibitemOpen
  \bibfield  {author} {\bibinfo {author} {\bibfnamefont {W.}~\bibnamefont
  {Wang}}, \bibinfo {author} {\bibfnamefont {T.}~\bibnamefont {Wang}}, \bibinfo
  {author} {\bibfnamefont {V.~P.}\ \bibnamefont {Amin}}, \bibinfo {author}
  {\bibfnamefont {Y.}~\bibnamefont {Wang}}, \bibinfo {author} {\bibfnamefont
  {A.}~\bibnamefont {Radhakrishnan}}, \bibinfo {author} {\bibfnamefont
  {A.}~\bibnamefont {Davidson}}, \bibinfo {author} {\bibfnamefont {S.~R.}\
  \bibnamefont {Allen}}, \bibinfo {author} {\bibfnamefont {T.~J.}\ \bibnamefont
  {Silva}}, \bibinfo {author} {\bibfnamefont {H.}~\bibnamefont {Ohldag}},
  \bibinfo {author} {\bibfnamefont {D.}~\bibnamefont {Balzar}}, \bibinfo
  {author} {\bibfnamefont {B.~L.}\ \bibnamefont {Zink}}, \bibinfo {author}
  {\bibfnamefont {P.~M.}\ \bibnamefont {Haney}}, \bibinfo {author}
  {\bibfnamefont {J.~Q.}\ \bibnamefont {Xiao}}, \bibinfo {author}
  {\bibfnamefont {D.~G.}\ \bibnamefont {Cahill}}, \bibinfo {author}
  {\bibfnamefont {V.~O.}\ \bibnamefont {Lorenz}},\ and\ \bibinfo {author}
  {\bibfnamefont {X.}~\bibnamefont {Fan}},\ }\bibfield  {title} {\bibinfo
  {title} {Anomalous spin--orbit torques in magnetic single-layer films},\
  }\href {https://doi.org/10.1038/s41565-019-0504-0} {\bibfield  {journal}
  {\bibinfo  {journal} {Nature Nanotechnology}\ }\textbf {\bibinfo {volume}
  {14}},\ \bibinfo {pages} {819} (\bibinfo {year} {2019})}\BibitemShut
  {NoStop}%
\bibitem [{\citenamefont {Céspedes-Berrocal}\ \emph
  {et~al.}(2021)\citenamefont {Céspedes-Berrocal}, \citenamefont {Damas},
  \citenamefont {Petit-Watelot}, \citenamefont {Maccariello}, \citenamefont
  {Tang}, \citenamefont {Arriola-Córdova}, \citenamefont {Vallobra},
  \citenamefont {Xu}, \citenamefont {Bello}, \citenamefont {Martin},
  \citenamefont {Migot}, \citenamefont {Ghanbaja}, \citenamefont {Zhang},
  \citenamefont {Hehn}, \citenamefont {Mangin}, \citenamefont {Panagopoulos},
  \citenamefont {Cros}, \citenamefont {Fert},\ and\ \citenamefont
  {Rojas-Sánchez}}]{Cespedes-Berrocal2021}%
  \BibitemOpen
  \bibfield  {author} {\bibinfo {author} {\bibfnamefont {D.}~\bibnamefont
  {Céspedes-Berrocal}}, \bibinfo {author} {\bibfnamefont {H.}~\bibnamefont
  {Damas}}, \bibinfo {author} {\bibfnamefont {S.}~\bibnamefont
  {Petit-Watelot}}, \bibinfo {author} {\bibfnamefont {D.}~\bibnamefont
  {Maccariello}}, \bibinfo {author} {\bibfnamefont {P.}~\bibnamefont {Tang}},
  \bibinfo {author} {\bibfnamefont {A.}~\bibnamefont {Arriola-Córdova}},
  \bibinfo {author} {\bibfnamefont {P.}~\bibnamefont {Vallobra}}, \bibinfo
  {author} {\bibfnamefont {Y.}~\bibnamefont {Xu}}, \bibinfo {author}
  {\bibfnamefont {J.-L.}\ \bibnamefont {Bello}}, \bibinfo {author}
  {\bibfnamefont {E.}~\bibnamefont {Martin}}, \bibinfo {author} {\bibfnamefont
  {S.}~\bibnamefont {Migot}}, \bibinfo {author} {\bibfnamefont
  {J.}~\bibnamefont {Ghanbaja}}, \bibinfo {author} {\bibfnamefont
  {S.}~\bibnamefont {Zhang}}, \bibinfo {author} {\bibfnamefont
  {M.}~\bibnamefont {Hehn}}, \bibinfo {author} {\bibfnamefont {S.}~\bibnamefont
  {Mangin}}, \bibinfo {author} {\bibfnamefont {C.}~\bibnamefont
  {Panagopoulos}}, \bibinfo {author} {\bibfnamefont {V.}~\bibnamefont {Cros}},
  \bibinfo {author} {\bibfnamefont {A.}~\bibnamefont {Fert}},\ and\ \bibinfo
  {author} {\bibfnamefont {J.-C.}\ \bibnamefont {Rojas-Sánchez}},\ }\bibfield
  {title} {\bibinfo {title} {{Current-Induced Spin Torques on Single GdFeCo
  Magnetic Layers}},\ }\href
  {https://doi.org/https://doi.org/10.1002/adma.202007047} {\bibfield
  {journal} {\bibinfo  {journal} {Advanced Materials}\ }\textbf {\bibinfo
  {volume} {33}},\ \bibinfo {pages} {2007047} (\bibinfo {year}
  {2021})}\BibitemShut {NoStop}%
\bibitem [{\citenamefont {M\"uller}\ \emph {et~al.}(2019)\citenamefont
  {M\"uller}, \citenamefont {Hoffmann}, \citenamefont {Di\ss{}elkamp},
  \citenamefont {Sch\"urhoff}, \citenamefont {Mavros}, \citenamefont
  {Sallermann}, \citenamefont {Kiselev}, \citenamefont {J\'onsson},\ and\
  \citenamefont {Bl\"ugel}}]{Mueller2019}%
  \BibitemOpen
  \bibfield  {author} {\bibinfo {author} {\bibfnamefont {G.~P.}\ \bibnamefont
  {M\"uller}}, \bibinfo {author} {\bibfnamefont {M.}~\bibnamefont {Hoffmann}},
  \bibinfo {author} {\bibfnamefont {C.}~\bibnamefont {Di\ss{}elkamp}}, \bibinfo
  {author} {\bibfnamefont {D.}~\bibnamefont {Sch\"urhoff}}, \bibinfo {author}
  {\bibfnamefont {S.}~\bibnamefont {Mavros}}, \bibinfo {author} {\bibfnamefont
  {M.}~\bibnamefont {Sallermann}}, \bibinfo {author} {\bibfnamefont {N.~S.}\
  \bibnamefont {Kiselev}}, \bibinfo {author} {\bibfnamefont {H.}~\bibnamefont
  {J\'onsson}},\ and\ \bibinfo {author} {\bibfnamefont {S.}~\bibnamefont
  {Bl\"ugel}},\ }\bibfield  {title} {\bibinfo {title} {Spirit: Multifunctional
  framework for atomistic spin simulations},\ }\href
  {https://doi.org/10.1103/PhysRevB.99.224414} {\bibfield  {journal} {\bibinfo
  {journal} {Phys. Rev. B}\ }\textbf {\bibinfo {volume} {99}},\ \bibinfo
  {pages} {224414} (\bibinfo {year} {2019})}\BibitemShut {NoStop}%
\bibitem [{\citenamefont {Dasgupta}\ and\ \citenamefont
  {Tchernyshyov}(2020)}]{Dasgupta2020}%
  \BibitemOpen
  \bibfield  {author} {\bibinfo {author} {\bibfnamefont {S.}~\bibnamefont
  {Dasgupta}}\ and\ \bibinfo {author} {\bibfnamefont {O.}~\bibnamefont
  {Tchernyshyov}},\ }\bibfield  {title} {\bibinfo {title} {{Theory of spin
  waves in a hexagonal antiferromagnet}},\ }\href
  {https://doi.org/10.1103/PhysRevB.102.144417} {\bibfield  {journal} {\bibinfo
   {journal} {Phys. Rev. B}\ }\textbf {\bibinfo {volume} {102}},\ \bibinfo
  {pages} {144417} (\bibinfo {year} {2020})}\BibitemShut {NoStop}%
\bibitem [{\citenamefont {Chen}\ \emph {et~al.}(2020)\citenamefont {Chen},
  \citenamefont {Gaudet}, \citenamefont {Dasgupta}, \citenamefont {Marcus},
  \citenamefont {Lin}, \citenamefont {Chen}, \citenamefont {Tomita},
  \citenamefont {Ikhlas}, \citenamefont {Zhao}, \citenamefont {Chen},
  \citenamefont {Stone}, \citenamefont {Tchernyshyov}, \citenamefont
  {Nakatsuji},\ and\ \citenamefont {Broholm}}]{Chen2020}%
  \BibitemOpen
  \bibfield  {author} {\bibinfo {author} {\bibfnamefont {Y.}~\bibnamefont
  {Chen}}, \bibinfo {author} {\bibfnamefont {J.}~\bibnamefont {Gaudet}},
  \bibinfo {author} {\bibfnamefont {S.}~\bibnamefont {Dasgupta}}, \bibinfo
  {author} {\bibfnamefont {G.~G.}\ \bibnamefont {Marcus}}, \bibinfo {author}
  {\bibfnamefont {J.}~\bibnamefont {Lin}}, \bibinfo {author} {\bibfnamefont
  {T.}~\bibnamefont {Chen}}, \bibinfo {author} {\bibfnamefont {T.}~\bibnamefont
  {Tomita}}, \bibinfo {author} {\bibfnamefont {M.}~\bibnamefont {Ikhlas}},
  \bibinfo {author} {\bibfnamefont {Y.}~\bibnamefont {Zhao}}, \bibinfo {author}
  {\bibfnamefont {W.~C.}\ \bibnamefont {Chen}}, \bibinfo {author}
  {\bibfnamefont {M.~B.}\ \bibnamefont {Stone}}, \bibinfo {author}
  {\bibfnamefont {O.}~\bibnamefont {Tchernyshyov}}, \bibinfo {author}
  {\bibfnamefont {S.}~\bibnamefont {Nakatsuji}},\ and\ \bibinfo {author}
  {\bibfnamefont {C.}~\bibnamefont {Broholm}},\ }\bibfield  {title} {\bibinfo
  {title} {{Antichiral spin order, its soft modes, and their hybridization with
  phonons in the topological semimetal ${\mathrm{Mn}}_{3}\mathrm{Ge}$}},\
  }\href {https://doi.org/10.1103/PhysRevB.102.054403} {\bibfield  {journal}
  {\bibinfo  {journal} {Phys. Rev. B}\ }\textbf {\bibinfo {volume} {102}},\
  \bibinfo {pages} {054403} (\bibinfo {year} {2020})}\BibitemShut {NoStop}%
\bibitem [{\citenamefont {Gomonaj}\ and\ \citenamefont
  {L'vov}(1992)}]{Gomonay1992}%
  \BibitemOpen
  \bibfield  {author} {\bibinfo {author} {\bibfnamefont {E.~V.}\ \bibnamefont
  {Gomonaj}}\ and\ \bibinfo {author} {\bibfnamefont {V.~A.}\ \bibnamefont
  {L'vov}},\ }\bibfield  {title} {\bibinfo {title} {Phenomenologic study of
  phase transitions in noncollinear antiferromagnets of metallic perovskite
  type},\ }\href {https://doi.org/10.1080/01411599208203457} {\bibfield
  {journal} {\bibinfo  {journal} {Phase Transitions}\ }\textbf {\bibinfo
  {volume} {38}},\ \bibinfo {pages} {15} (\bibinfo {year} {1992})}\BibitemShut
  {NoStop}%
\bibitem [{\citenamefont {Suzuki}\ \emph {et~al.}(2017)\citenamefont {Suzuki},
  \citenamefont {Koretsune}, \citenamefont {Ochi},\ and\ \citenamefont
  {Arita}}]{Suzuki2017}%
  \BibitemOpen
  \bibfield  {author} {\bibinfo {author} {\bibfnamefont {M.-T.}\ \bibnamefont
  {Suzuki}}, \bibinfo {author} {\bibfnamefont {T.}~\bibnamefont {Koretsune}},
  \bibinfo {author} {\bibfnamefont {M.}~\bibnamefont {Ochi}},\ and\ \bibinfo
  {author} {\bibfnamefont {R.}~\bibnamefont {Arita}},\ }\bibfield  {title}
  {\bibinfo {title} {{Cluster multipole theory for anomalous Hall effect in
  antiferromagnets}},\ }\href {https://doi.org/10.1103/PhysRevB.95.094406}
  {\bibfield  {journal} {\bibinfo  {journal} {Phys. Rev. B}\ }\textbf {\bibinfo
  {volume} {95}},\ \bibinfo {pages} {094406} (\bibinfo {year}
  {2017})}\BibitemShut {NoStop}%
\bibitem [{\citenamefont {Nomoto}\ and\ \citenamefont
  {Arita}(2020)}]{Nomoto2020}%
  \BibitemOpen
  \bibfield  {author} {\bibinfo {author} {\bibfnamefont {T.}~\bibnamefont
  {Nomoto}}\ and\ \bibinfo {author} {\bibfnamefont {R.}~\bibnamefont {Arita}},\
  }\bibfield  {title} {\bibinfo {title} {Cluster multipole dynamics in
  noncollinear antiferromagnets},\ }\href
  {https://doi.org/10.1103/PhysRevResearch.2.012045} {\bibfield  {journal}
  {\bibinfo  {journal} {Phys. Rev. Research}\ }\textbf {\bibinfo {volume}
  {2}},\ \bibinfo {pages} {012045(R)} (\bibinfo {year} {2020})}\BibitemShut
  {NoStop}%
\bibitem [{\citenamefont {Šmejkal}\ \emph {et~al.}(2021)\citenamefont
  {Šmejkal}, \citenamefont {Sinova},\ and\ \citenamefont
  {Jungwirth}}]{Smejkal2021}%
  \BibitemOpen
  \bibfield  {author} {\bibinfo {author} {\bibfnamefont {L.}~\bibnamefont
  {Šmejkal}}, \bibinfo {author} {\bibfnamefont {J.}~\bibnamefont {Sinova}},\
  and\ \bibinfo {author} {\bibfnamefont {T.}~\bibnamefont {Jungwirth}},\
  }\href@noop {} {\bibinfo {title} {Altermagnetism: spin-momentum locked phase
  protected by non-relativistic symmetries}} (\bibinfo {year} {2021}),\ \Eprint
  {https://arxiv.org/abs/arXiv:2105.05820} {arXiv:2105.05820} \BibitemShut
  {NoStop}%
\end{thebibliography}%


%

\end{document}